\newif\ifcomm
\commfalse

\newif\ifblind
\blindfalse

\newif\ifconf
\conftrue
\newif\ifacm
\acmtrue

\newif\ifacmart
\ifacm
   \acmarttrue
\else
   \acmartfalse
\fi

\newif\ifs
\sfalse

\ifconf
    \newcommand{\Conf}[1]{#1}
    \newcommand{\TR}[1]{}
    \newcommand{\Journal}[1]{}  
    \newcommand{\OnlyTR}[1]{}   
\else
    \newcommand{\Conf}[1]{}
    \newcommand{\TR}[1]{#1}
    \newcommand{\Journal}[1]{}  
    \newcommand{\OnlyTR}[1]{#1}   
\fi

\documentclass[newfonts=false,format=sigconf,10pt,letterpaper]{acmart}
\PassOptionsToPackage{numbers,super}{natbib}

\usepackage{times}  

\hypersetup{pdfstartview=FitH,pdfpagelayout=SinglePage}
\usepackage{hyperref}
\definecolor{darkred}{rgb}{0.7,0,0}
\definecolor{darkgreen}{rgb}{0,0.5,0}
\hypersetup{colorlinks=true,
        linkcolor=darkred,
        citecolor=darkgreen}
\ifs
     
    \ifacmart
        \usepackage[small,bf]{caption}
    \else
           \usepackage[hang,small,bf]{caption}
    \fi
\else
    
\fi

\usepackage[utf8]{inputenc}
\usepackage[inline]{enumitem}
\newlist{inlinelist}{enumerate*}{1}
\setlist[inlinelist]{label=(\arabic*)}
\newlist{romanlist}{enumerate*}{1}
\setlist[romanlist]{label=(\roman*)}

\usepackage[english]{babel}
\usepackage{times}
\usepackage{amsmath}
\usepackage{setspace}

\usepackage{amsfonts}
\usepackage{graphicx}
\usepackage{caption}
\usepackage{subcaption}
\usepackage{textcomp}
\usepackage{xspace}

\ifcomm
	\newcommand{\mycomm}[3]{{\color{#2} \textbf{[#1: #3]}}} 
\else
    \newcommand{\mycomm}[3]{}
\fi

\providecommand{\vs}{vs. }
\providecommand{\ie}{\emph{i.e.,} }
\providecommand{\eg}{\emph{e.g.,} }

\providecommand{\etc}{\emph{etc.} }

\def\compactify{\itemsep=0pt \topsep=0pt \partopsep=0pt \parsep=0pt}
  \let\latexusecounter=\usecounter

\newcommand{\T}[1]{\smallskip\noindent\textbf{#1}} 
\newcommand{\oursys}{Pied Piper\xspace}
\newcommand{\name}{Pied Piper\xspace}
\newcommand{\ksplit}{K-Split\xspace}
\newcommand{\sockets}{Berkeley sockets\xspace}
\newcommand{\proxies}{relays\xspace}
\newcommand{\reconn}{reusable connections\xspace}
\newcommand{\usec}{$\mu$sec\xspace}
\newcommand{\newVar}[2]{\newcommand{#1}{\ensuremath{#2}\xspace}}
\newVar{\rc}{R_C}
\newVar{\rs}{R_S}
\newVar{\rmid}{R_M}
\newVar{\rtt}{T}

\title{Pied Piper: Rethinking Internet Data Delivery}
\newcommand{\aut}[2]{#1\texorpdfstring{$^{#2}$}{(#2)}} 
\author{
  \aut{Aran Bergman}{1,2},
  \aut{Israel Cidon}{1},
  \aut{Isaac Keslassy}{1,2},
  \aut{Noga Rotman}{3},
  \aut{Michael Schapira}{3},
  \aut{Alex Markuze}{1,2},
  \aut{Eyal Zohar}{1}
}%
      \affiliation{
$^1$ \textit{VMware} \quad 
$^2$ \textit{Technion} \quad 
$^3$ \textit{HUJI} \quad 
	}
\date{}

\ifacm 
 \ifacmart
 \acmConference[SIGCOMM'18]{ACM Conference}{August 21-23, 2018}{Budapest, Hungary}
    \setcopyright{none}
	\renewcommand\footnotetextcopyrightpermission[1]{} 
 \fi
\fi

\begin{document}

 \sloppypar

\begin{abstract}

We contend that, analogously to the transition from resource-limited on-prem computing to resource-abundant cloud computing, Internet data delivery should also be adapted to a reality in which the cloud offers a virtually unlimited resource, \ie network capacity, and virtualization enables delegating local tasks, such as routing and congestion control, to the cloud. This necessitates rethinking the traditional roles of  inter- and intra-domain routing and conventional end-to-end congestion control.

We introduce \textit{Optimized Cloudified Delivery (OCD)}, a holistic approach for optimizing joint Internet/cloud data delivery, and evaluate OCD through hundreds of thousands of file downloads from multiple locations. We start by examining an OCD baseline approach:  traffic from a source $A$ to a destination $B$ successively passes through two cloud virtual machines operating as relays - nearest to $A$ and $B$; and the two cloud relays employ TCP split.

We show that even this na\"ive strategy can outperform recently proposed improved end-to-end congestion control paradigms (BBR and PCC) by an order of magnitude.

Next, we present a protocol-free, ideal pipe model of data transmission, 
and identify where today's Internet data delivery mechanisms diverge from this model. We then design and implement OCD \textit{\name}. \name leverages various techniques, including novel kernel-based transport-layer accelerations, to improve the Internet-Cloud interface so as to approximately match the ideal network pipe model.
\end{abstract}


\maketitle

\section{Introduction}

{\let\thefootnote\relax\footnotetext{
This work has been submitted to the IEEE for possible publication. Copyright may be transferred without notice, after which this version may no longer be accessible.
}}

The Internet's data delivery infrastructure, whose core components (BGP, TCP, \textit{etc.}) were designed decades ago, reflects a traditional view of the Internet's landscape: traffic leaving a source traverses multiple organizational networks (Autonomous Systems), potentially contending over network resources with other traffic flows en route to its destination at multiple locations along the way. Indeed, despite many changes to the Internet ecosystem over the past two decades (the emergence of global corporate WANs, IXPs, CDNs), BGP paths consistently remain about 4-AS-hop-long on average~\cite{kuhne2012ripe,apnic2017bgp}, and congestion control on the Internet remains notoriously bad~\cite{PCC}. Consequently, fixing the deficiencies of the Internet's core networking protocols for routing and congestion control, though a decades-old agenda, remains at the center of attention, as evidenced by the surge of recent interest in novel congestion control paradigms~\cite{PCC,BBR,keqiang2016acdc} and in data-driven BGP-path-selection~\cite{yap2017taking,schlinker2017engineering}
.

Meanwhile, recent developments give rise to new opportunities. Major cloud providers are pouring billions of dollars into high-performance globe-spanning networking infrastructures~\cite{gcp2018spending,aws2018spending,microsoft2018spending}. In addition, virtualization enables setting up logical relays within the cloud (in the form of virtual machines (VMs)), and controlling the routing and transport-layer protocols executed by these relays. As a result, any user can establish a virtually infinite number of VMs in the cloud, each potentially sending at a virtually infinite capacity to its outgoing links (\eg, up to 2Gbps per VM in our measurements on the most basic VM types), in a pay-as-you-go model where rates per GB keep falling. 

We argue that the emergence of cloud networks of this scale and capacity compels the networking community to rethink the traditional roles of  inter- and intra-domain routing, and of conventional end-to-end congestion control. In fact, we contend that \textit{networking is following the path of computing and storage} in their transition from resource-limited on-prem computing to resource-abundant cloud computing.  Similarly to storage and computing, in the context of networking, too, the cloud offers a virtually unlimited resource, \ie network capacity, and virtualization enables delegating to the cloud local tasks, namely, (interdomain) routing and congestion control. Our aim is to analyze what would happen to Internet data delivery if it also were to become ``cloudified''.

\T{OCD baseline (\S\ref{sec:ocd-baseline}).} We introduce \textit{Optimized and Cloudified Delivery}~(OCD), a holistic approach that aims at exploring the performance  implications  of delegating routing and congestion control to the cloud. We start by comparing the current end-to-end Internet data delivery to an \textit{OCD baseline}: (1) traffic from a source $A$ to a destination $B$ successively passes through the two cloud relays, one near $A$ and the other near $B$; and (2) the two cloud relays employ TCP split, dividing the end-to-end TCP connection into three connections. Thanks to the advent of virtualization, this OCD baseline is relatively easy to implement. This baseline stategy, while simple, captures the essence of OCD: minimizing time outside the cloud so as to avoid BGP's inefficiencies and separating traffic modulation within and without the cloud.

Our evaluation of the OCD baseline on three major clouds and on both inter- and intra-continental traffic, reveals that file download times for a 4MB file over this OCD baseline outperform those over the wild Internet by a $5\times$ factor. Download times of large files can improve by up to $30\times$, while those of small files are barely affected or get slightly worse. Importantly, using recently proposed end-to-end congestion-control protocols over the Internet, such as PCC~\cite{PCC} and BBR~\cite{BBR}, improves performance by less than $30\%$. In short, the OCD baseline approach appears quite promising. 

\T{\name (\S\ref{sec:rate-control}).} To go beyond the OCD baseline, we next introduce the abstraction of a fixed-capacity network pipe interconnecting source $A$ and destination $B$. We present a protocol-free, ideal theoretical model of such a network pipe, and explain how today's Internet data delivery mechanisms introduce a seriality that diverges from this ideal pipe model. We argue that as the (long) cloud leg essentially encounters no congestion, transport-layer mechanisms can be adapted to better approximate the ideal network pipe model. To accomplish this, we design and implement OCD \textit{\name}, which leverages various techniques, including novel kernel-based transport-layer accelerations. 

\T{OCD Evaluations (\S\ref{sec:baseline-good_enough}).} We next revisit two key design choices behind our OCD approach: (1) employing two cloud relays (placed near the source and the destination) and (2) splitting TCP. Our evaluation relies on a large-scale empirical
investigation involving the download of hundreds of thousands of files of
varying sizes from multiple locations within and outside the cloud. Our results
show that: (1) Using the two cloud relays that are closest to $A$ and $B$ indeed improves performance over using most single cloud relays. While using the best single cloud relay may perform slightly better, it is hard to identify this relay in advance. Further, using three relays instead of two provides, at best, small marginal benefit. Also, (2) not splitting TCP significantly hurts the data transfer performance, and therefore splitting appears to be a \textit{must}. In addition, we find that combining routes through several clouds may help in the future.

\T{OCD implementation (\S\ref{sec:implementation}).} The OCD \name relies on a novel Linux kernel-based split acceleration we call \textit{\ksplit}~\cite{ktcp}. To our knowledge, \ksplit is the first TCP split tool written in the Linux kernel, using kernel sockets. Avoiding the penalties that stem from the numerous system calls of user-mode implementations. We make \ksplit publicly available as open-source on Github, thus enabling any user to try \name on any cloud. 

\T{Deployment strategies (\S\ref{sec:deployment}).} We finally discuss the issues arising from a large global transition to OCD. A first danger is the saturation of cloud resources. However, since these resources are based on a pay-as-you-go model, we posit that the invisible hand will incentivize cloud providers to provision their networks to meet the growing user demand.  
A second danger is the potential unfairness between high- and low-speed data transfer lanes. Paradoxically, we show that OCD may be a way to \textit{restore} fairness between long- and short-span flows that are obeying the unfairness of TCP mechanisms (bandwidth~$\propto \frac{1}{RTT}$).

\begin{figure*}[t]
  \centering
  \begin{subfigure}{.49\textwidth}
  \centering
    \includegraphics[width=0.98\textwidth,clip=true, trim = 0 115mm 0 0]{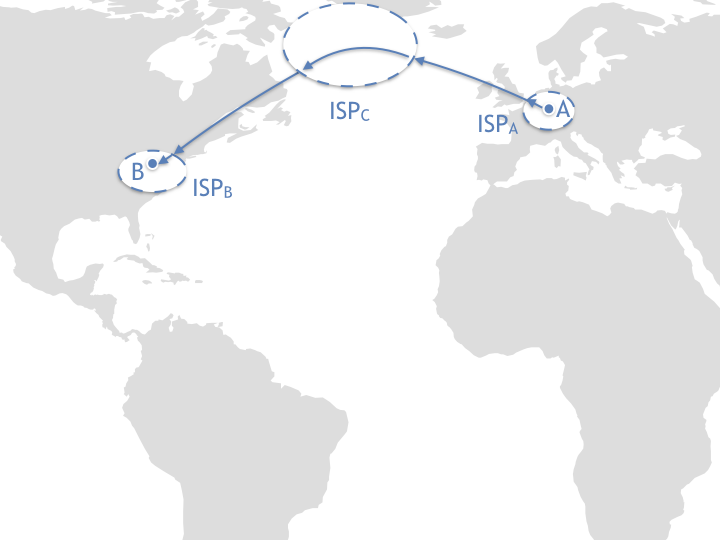}
    \caption{\textit{e2e}: Default  transfer through the Internet}
    \label{fig:e2e-traffic}
\end{subfigure}
\begin{subfigure}{.49\textwidth}
  \centering
    \includegraphics[width=0.98\textwidth, clip, trim = 0 115mm 0 0]{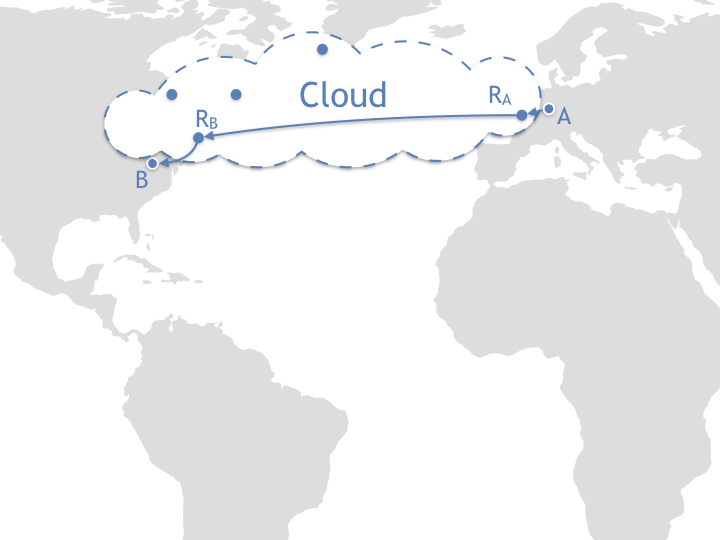}
    \caption{\textit{OCD Baseline}: transfer through the Internet and the cloud}
    \label{fig:cloud-traffic}
\end{subfigure}
\caption{\textit{e2e} (end-to-end) \vs \textit{OCD Baseline}: Data transfer from Server A in Frankfurt, Germany to Client B in New York, NY, US 
}
\label{fig:e2e-vs-cloud-traffic}
\end{figure*}

\section{OCD Baseline
}\label{sec:ocd-baseline}

We present below a baseline proposal for OCD. We show that even the proposed na\"ive solution significantly outperforms utilizing recently proposed improved end-to-end congestion control schemes.

\subsection{Baseline Strategy}

Consider an end-host $A$ in Frankfurt that is sending traffic to an end-host $B$ in New~York. We will refer to $A$ as the server and $B$ as the client. Under today's default data transport mechanisms, traffic from $A$ to $B$ will be routed along the BGP path between the two. As illustrated in Figure~\ref{fig:e2e-traffic}, such a path might connect $A$'s ISP to $B$'s ISP via an intermediate ISP (or more). The modulation of traffic from $A$ to $B$ in our example is via TCP congestion control, which reacts to network conditions on the path (congestion, freed capacity) by adaptively adjusting the sending rate.        

We consider the following simple strategy for the \textit{cloudified} data delivery,
as illustrated in Figure~\ref{fig:cloud-traffic}: Traffic from $A$ is sent
directly to a geographically-close cloud ingress point. Then, traffic traverses the (single) cloud's infrastructure and leaves that cloud en
route to $B$ at an egress point close to $B$. In addition, the
transport-layer connection between $A$ and $B$ is split into $3$ separate connections, each employing TCP (Cubic) congestion
control: between $A$ and the cloud, within the cloud, and between the cloud and
$B$. We point out that this strategy does not incorporate aspects
such as employing other rate control schemes, leveraging additional relays
within the cloud, traversing multiple clouds, and more. Yet, as our empirical
results below demonstrate, even the simple baseline strategy yields significant
performance benefits. We later revisit the rate-control and routing aspects of the baseline approach in Sections~\ref{sec:rate-control} and~\ref{subsec:num_of_relays}, respectively.
To conclude, the baseline strategy relies on:

\vspace{0.1in}\noindent{\bf Two cloud relays, one geographically-close to the server and one geographically-close to the client.} We set up, for each client-server pair, two cloud relays (VMs), selected according to their geographical proximity to the end-point (the first closest to the server and the second closest to the client). Naturally, if the server/client are already in the cloud, then the corresponding (close) cloud relay(s) are not needed. The simple strategy does not utilize any more cloud relays between the ingress and egress points, nor traverses more than a single cloud.

\vspace{0.1in}\noindent{\bf Splitting the transport-layer connection.} We break the end-to-end connection between the two end-hosts into $3$ consecutive connections. As discussed in Section~\ref{sec:rate-control}, the challenges facing rate-control for the intra-cloud connection are fundamentally different than those faced by rate-control for the first and last connections (to/from the cloud). The baseline strategy ignores this distinction and simply employs traditional TCP congestion control independently on each leg.

\vspace{0.1in}\noindent
In~\ref{subsec:ocd-baseline_results} we contrast the baseline OCD strategy with other natural strategies: (1)  employing recent proposals for end-to-end congestion control (namely PCC~\cite{PCC} and BBR~\cite{BBR}), (2) employing TCP CUBIC with large initial cwnd, initial rwnd, and socket buffer,
and (3) routing traffic via the same cloud relays without splitting the transport-layer connection. Our results show that the OCD baseline strategy significantly outperforms all three strategies, suggesting that replacing the default congestion control alone, or the default routing alone, is insufficient.

\subsection{Experimental Framework}

\vspace{0.1in}\noindent{\bf Locations.} 
We present results from experiments performed on three different routes; (1)~India to US West Coast, (2)~Europe to US East Coast, and (3)~US East Coast to West Coast.

\vspace{0.1in}\noindent{\bf Clients and servers.} We used two types of clients: (1)~a personal computer located in San Francisco (SF), connected to the Internet through Comcast; (2)~PlanetLab~\cite{PlanetLab} nodes in multiple locations worldwide.
For servers outside the cloud, we used the following:
(1)~www.cc.iitb.ac.in, located at IIT Bombay, India; (2)~www.college.columbia.edu, located at Columbia University, NY, US;
(3)~Virtual machines on Kamatera~\cite{kamatera} infrastructure, located in NY, US, and Amsterdam, The Netherlands, using the cheapest settings available. 
We chose university-based servers as these typically have fairly good Internet connection but do not use CDNs, and so download times are not expected to be affected by redirections or caching. 

\vspace{0.1in}\noindent{\bf Downloaded files.} We downloaded two types of files from these servers: large files (3.9~MB for the first server, 3.8~MB for the second, 10~MB for the third type), and small files (17~KB for the first server, 18~KB for the second, 10~KB for the third type). Using Kamatera servers, we experimented with additional file sizes: 100~KB, 1~MB and 100~MB. 

\vspace{0.1in}\noindent{\bf Cloud relays.} To assess cloud performance, we deployed relays (virtual machines) on three major clouds: \textit{AWS} (Amazon Web Services), \textit{GCP} (Google Cloud Platform) and Microsoft \textit{Azure}. In each cloud, we deployed at least one relay in each publicly-available region (\autoref{tab:cloud-config}), 
\begin{table} {\small
    \centering 
    \begin{tabular}{c c c c}
         Cloud &            AWS &           Azure               & GCP \\ \hline 
         \# of Regions &    14 &            26                  & 10 \\
         Machine 
            &     t2.micro &      Standard\_DS1\_v2   & n1-standard-1 \\
         \textcent/hour &   1.2 -- 2         & 7                 & 4.75 -- 6.74 \\
         \textcent/GB & 9  & 8.7 -- 18.1   & 12 -- 23 (1 in US) \\ \hline
       
    \end{tabular}
    \caption{Number of regions and machine types used in each cloud, and their pricing, taken from the cloud providers' website. 
}
    \label{tab:cloud-config}
    }
\end{table}
yielding a total of $50$ potential relays. Each relay ran either Ubuntu 17.04 or 17.10 using relatively cheap machine types.

\vspace{0.1in}\noindent{\bf Routing through relays.} Routing through cloud relays without splitting the connection was performed by using NAT on the cloud machines, configured using Linux's iptables \cite{iptables}. NAT is used so that the return path traverses the same relays.

\vspace{0.1in}\noindent{\bf Splitting the connection.} To split the TCP connection we used ssh to localhost on each relay and utilized ssh's port forwarding option to forward the byte stream to the next machine en route to the destination.

\vspace{0.1in}\noindent{\bf Tools and techniques.} \textit{RTT measurements} are conducted using hping3 \cite{hping3}, by sending 20 SYN packets (at 0.1 seconds intervals) to an open TCP port on the other side and measuring the round-trip-time between sending each SYN and getting its corresponding SYN-ACK. The minimum of these 20 measurements is taken as the RTT between the two tested endpoints. We chose this method over ping so as to avoid the risk of ICMP filtering.

\vspace{0.1in}\noindent{\bf Congestion control protocols.} For testing different congestion control protocols, we used the Ubuntu implementation of BBR, and an in-house PCC kernel module. In addition, we tested TCP Cubic with large initial cwnd, initial rwnd, and socket buffer, an approach we term  \textit{``Turbo-Start Cubic''}. 

\vspace{0.1in} We ran all the experiments for a total of $8$ weeks, 
at different times of day. We repeated each of the above-described experiments at least 50 times and computed averaged results.\footnote{Some of the scripts used for this experimentation are made available through an anonymous github repository.}

\subsection{Results}\label{subsec:ocd-baseline_results}

\begin{figure}[t]
  \centering
  \includegraphics[width=0.9\columnwidth,trim=2mm 2mm 2mm 2mm,clip]{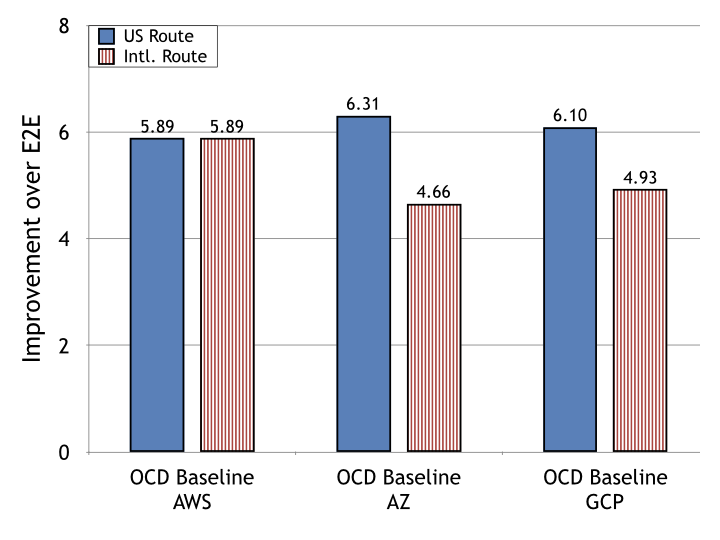}
  \caption{Evaluating OCD Baseline compared to e2e. We consider three major public clouds (Amazon AWS, Microsoft Azure, Google GCP) and two different routes: (1)~intra-US, where the server is located in the East Coast and transfers data to a residential client in the West Coast, and (2)~international route, where the server is located in India and sends data to the same residential client in the West Coast. The file size in both cases is about 4MB. OCD baseline provides a noticeable improvement over the default data delivery across all clouds.
  }
 \label{fig:ocd_baseline_initial}
\end{figure}

\vspace{0.1in}\noindent{\bf Baseline \vs Internet e2e.} \autoref{fig:ocd_baseline_initial} demonstrates the gains of our OCD Baseline strategy over default data transfer via the Internet. The settings used in these experiments are close to a realistic scenario, as in this case we utilize a residential client with good connectivity to the Internet, and web servers that are not in the cloud and do not utilize CDNs.

\begin{figure*}[t]
  \centering
  \includegraphics[width=0.98\textwidth,trim=5mm 0mm 0mm 0mm,clip]{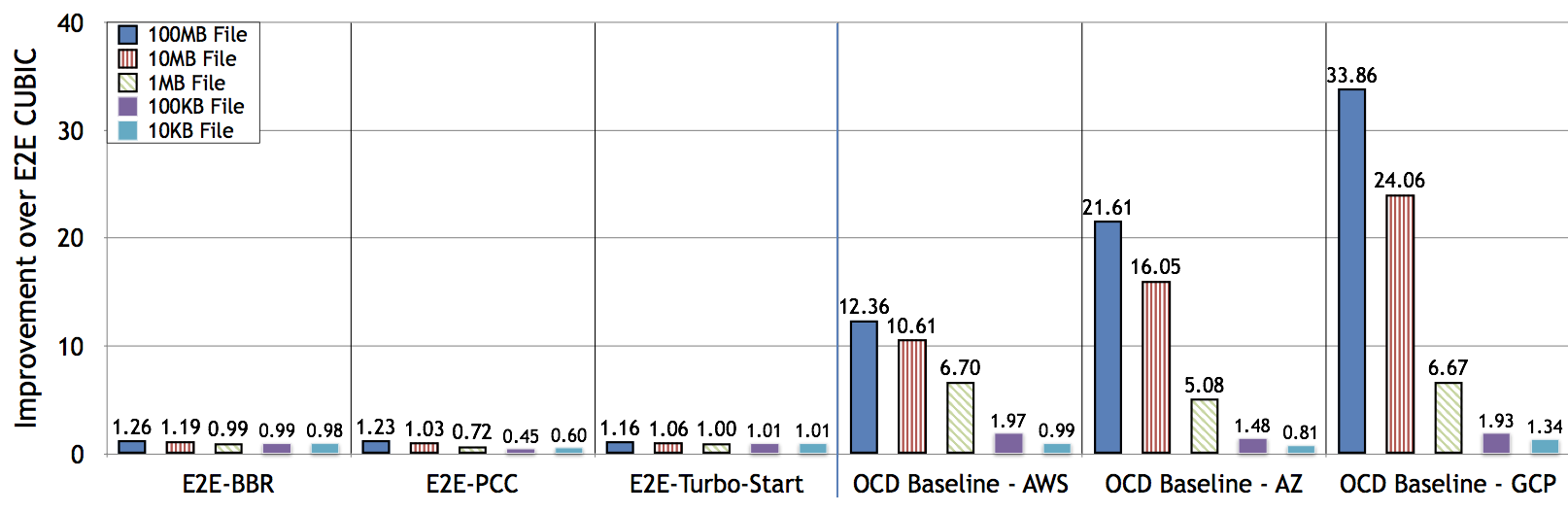}
  \caption{Clever congestion control strategies vs. OCD Baseline, for the international route between a Kamatera server in Amsterdam and a PlanetLab client in the East Coast. The improvement for large (10~MB) and extra-large (100~MB) is quite staggering, peaking at an improvement rate of $33\times$ for the extra-large file using GCP as the cloud provider. Note that as the file size decreases, the improvement declines as well, to the point where the OCD baseline might actually hurt the performance for smallest file size tested. We try to improve upon the OCD baseline for that end, amongst others, in Section~\ref{sec:rate-control}.}
 \label{fig:cloud-vs-CC}
\end{figure*}

\vspace{0.1in}\noindent{\bf Baseline \vs improved e2e congestion control.} We tested the OCD Baseline strategy against several  congestion control schemes  using Kamatera servers (which enabled us to change the congestion control scheme used by the server). \autoref{fig:cloud-vs-CC} showcases the results for a variety of file sizes, ranging from very small (10~KB) to extra-large (100~MB), using the second international route (Europe to East Coast).
While the improvement rates vary from cloud to cloud, all improve upon the default data transfer by a factor of at least $10\times$ for the large files, and at least $12\times$ for the extra-large files. Employing different congestion control schemes hardly made an impact, gaining a maximum of $1.26\times$ improvement.

\begin{figure}[h]
  \centering
  \includegraphics[width=\columnwidth,trim=2mm 2mm 2mm 2mm,clip]{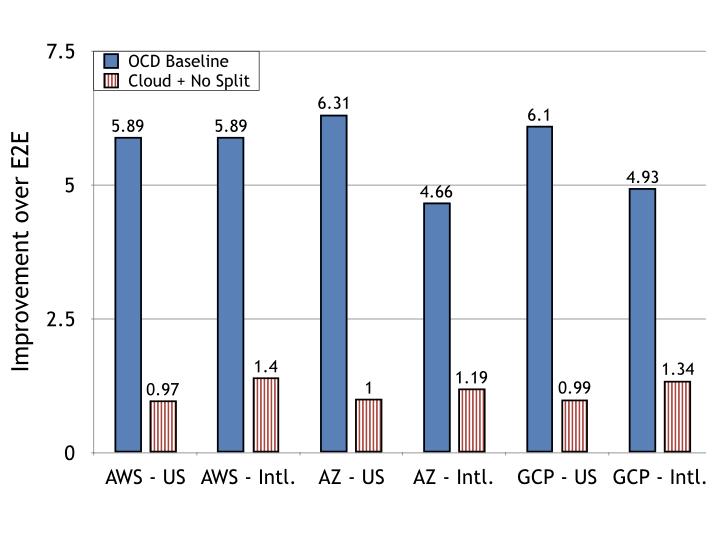}
  \caption{Routing through the cloud is not enough; in order to achieve improvement, one \textit{must} split the TCP connection. The same settings as in \autoref{fig:ocd_baseline_initial} are used.
  }
 \label{fig:must-split}
\end{figure}

\vspace{0.1in}\noindent{\bf Baseline (with splitting) \vs no splitting.} 
We validate the choice of TCP splitting via exhaustive testing.~\autoref{fig:must-split}, elaborating on \autoref{fig:ocd_baseline_initial}, illustrates our results. Similar gains occurred in other settings as well. The conclusion is clear: in this framework, splitting seems to be a \textit{must}.

\section{Better OCD with Pied Piper}\label{sec:rate-control}

\subsection{The Dream Network Pipe}\label{subsec:dream-pipe}

\begin{figure*}[!t]
  \centering
    \begin{subfigure}{0.65\columnwidth}
  \centering
  \includegraphics[width=\columnwidth]{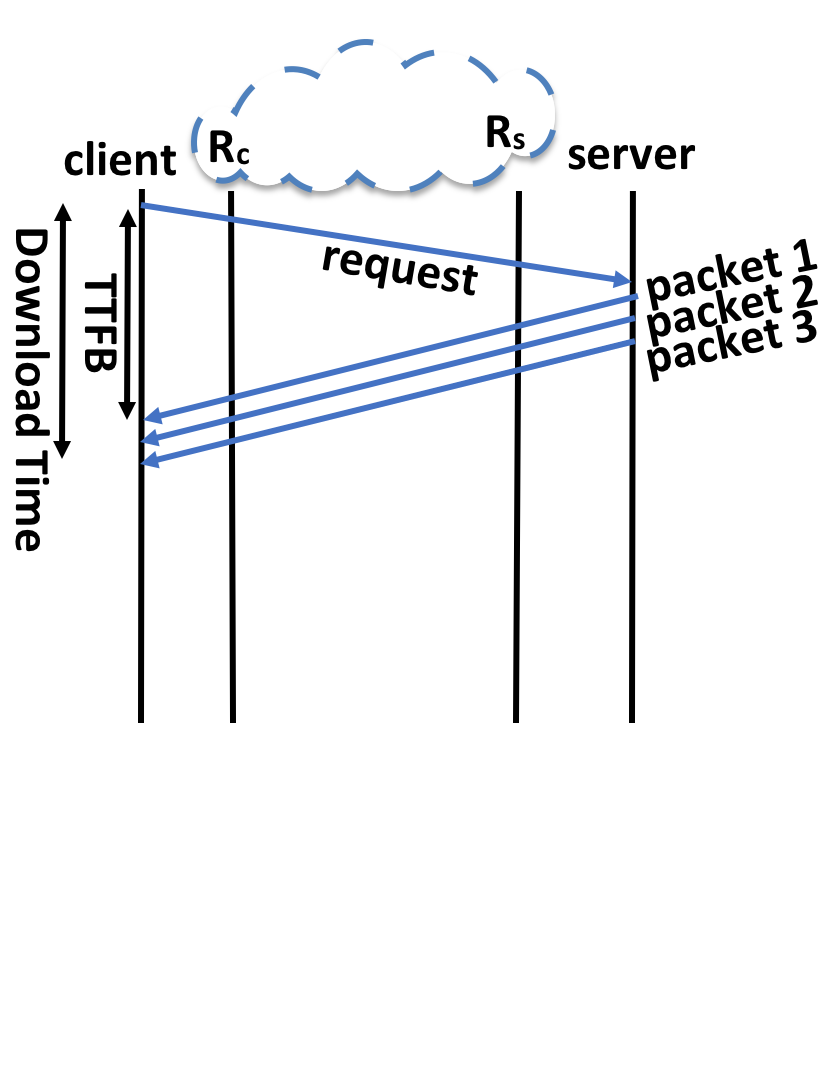}
    \caption{Ideal protocol-free transmission.}
    \label{fig:ideal}
\end{subfigure}    \centering
\begin{subfigure}{0.65\columnwidth}
  \centering
  \includegraphics[width=\columnwidth]{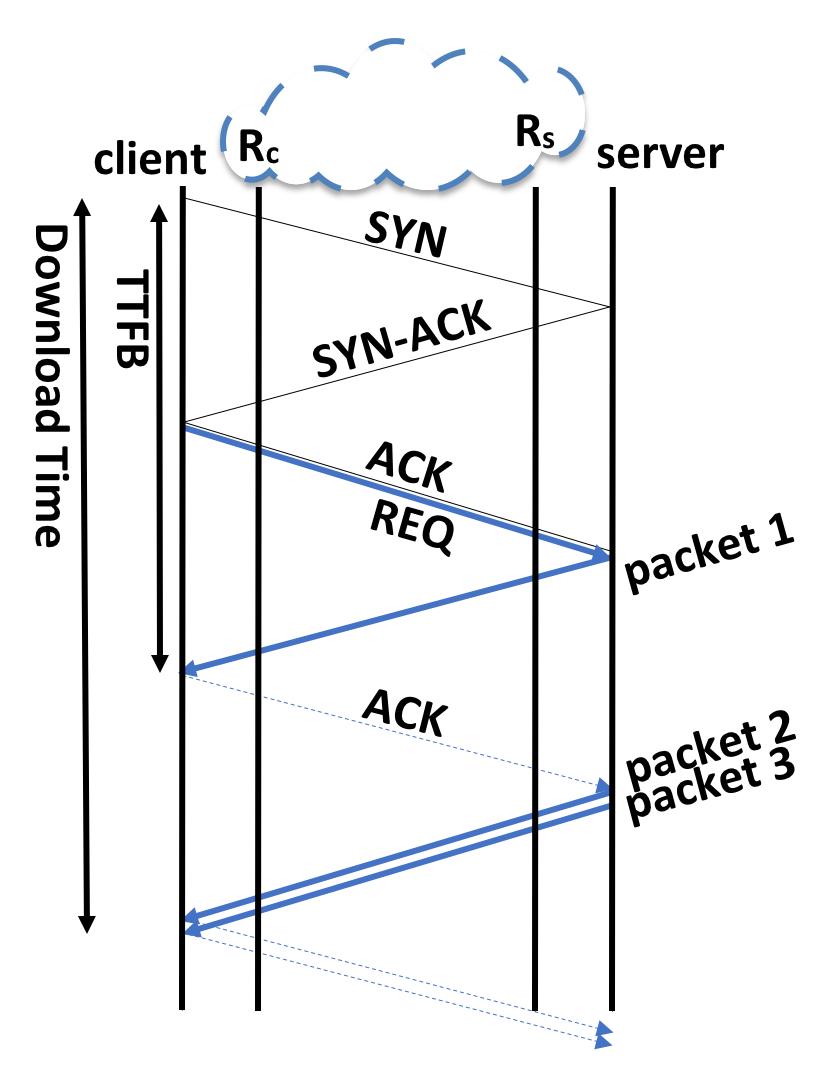}
    \caption{End-to-end through the Internet (or cloud).} \label{fig:e2e}
\end{subfigure}    \centering
\begin{subfigure}{0.65\columnwidth}
  \centering
  \includegraphics[width=\columnwidth,clip]{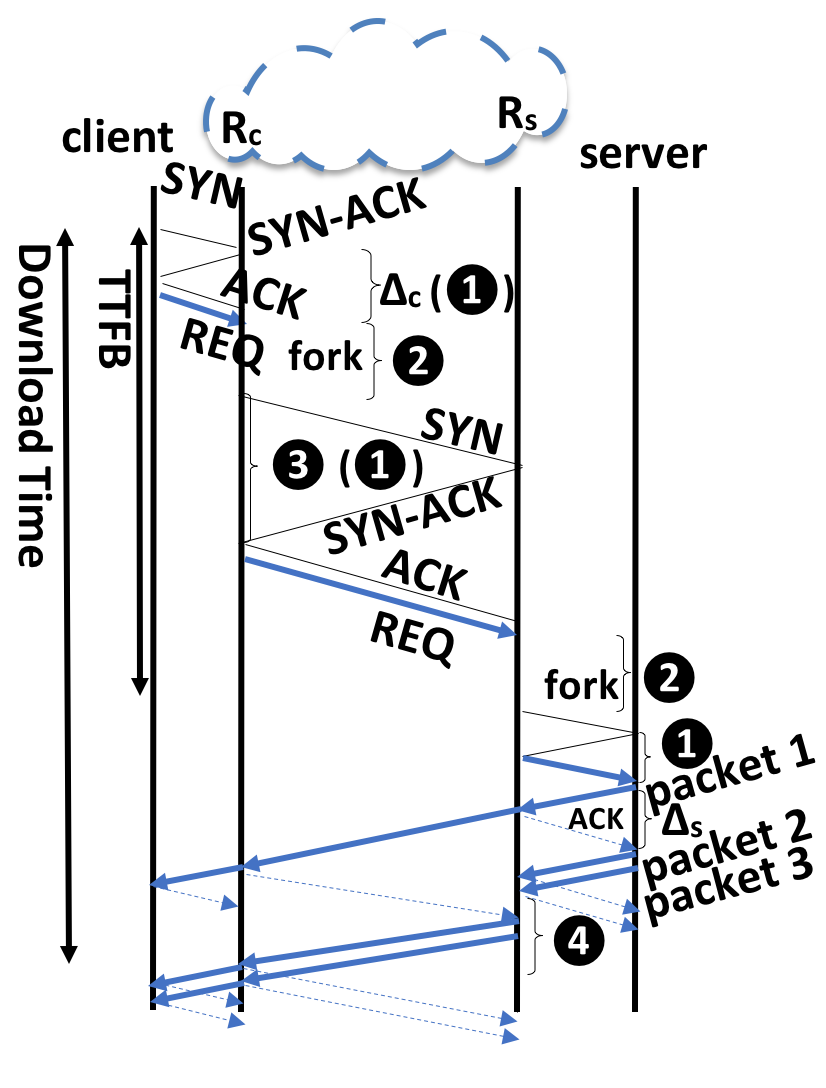}
    \caption{Simple TCP cloud split.} \label{fig:baseline}
\end{subfigure}
    \caption{Illustrated comparison of the considered baseline data transmission methods. Suppose, for simplicity of exposition, that the client requests three MSS-sized packets using HTTP over TCP, that the initial TCP window size is one MSS, and that there are no losses.\\ \textit{(a)} In an ideal clean-slate world, the request for packets would go through directly, triggering the transmission of all response packets. The time-to-first-byte (TTFB) is just one round-trip time (RTT), the lowest possible TTFB. The download time is barely higher.\\ \textit{(b)} End-to-end TCP transmission first requires the establishment of an end-to-end connection, adding one RTT to the ideal finish time. Waiting one RTT for the first ACK further delays the download. \\ \textit{(c)} The baseline strategy of Section~\ref{sec:ocd-baseline} decreases the ACK times for longer files, but introduces new delays such as thread fork delays in the connection. This explains why the Baseline is less efficient for small files. We later revisit all the delays that appear beyond those of the ideal transmission (Figure~\ref{fig:KSPLIT-improvements}). We show how to address the delays marked (1)-(4), but are left with delays $\Delta_c$ and $\Delta_s$ on the client and server sides, respectively.  }
\end{figure*}

\begin{figure*}[!t]
  \centering
    \begin{subfigure}{0.48\columnwidth}
  \centering
  \includegraphics[width=\columnwidth]{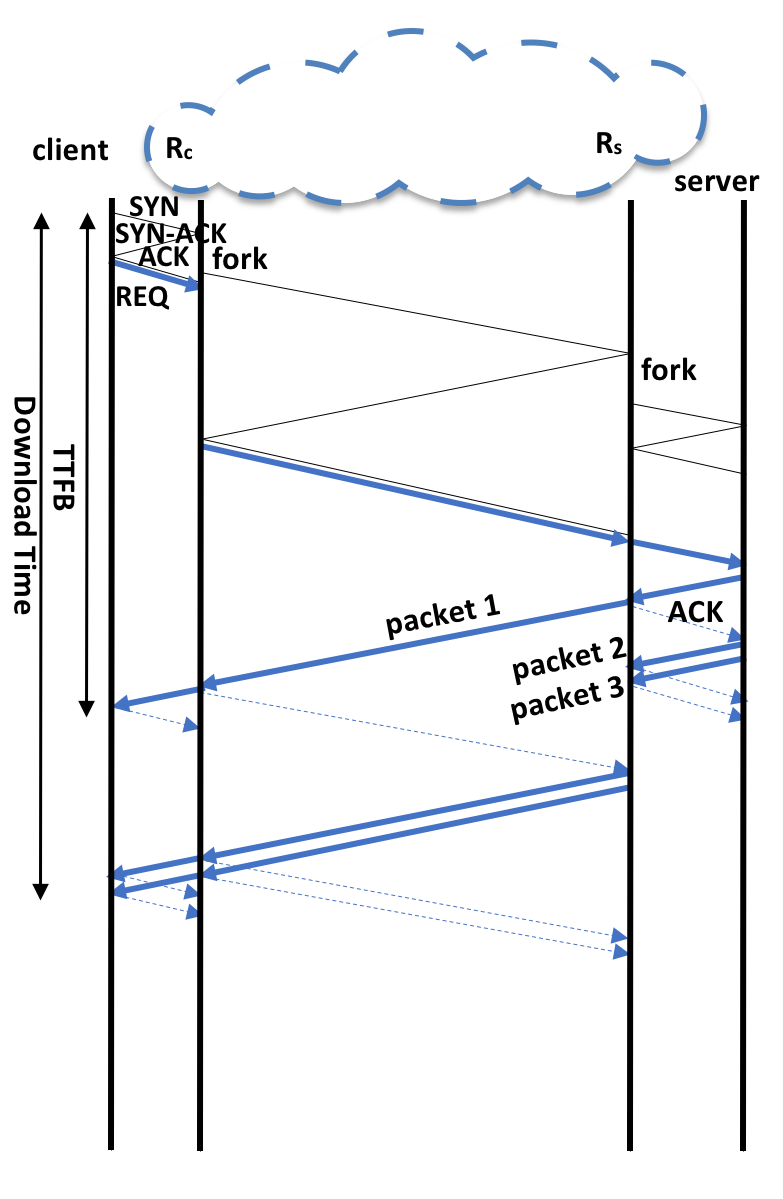}
    \caption{Early-SYN.}
    \label{fig:early-syn}
\end{subfigure}    \centering
\begin{subfigure}{0.48\columnwidth}
  \centering
  \includegraphics[width=\columnwidth]{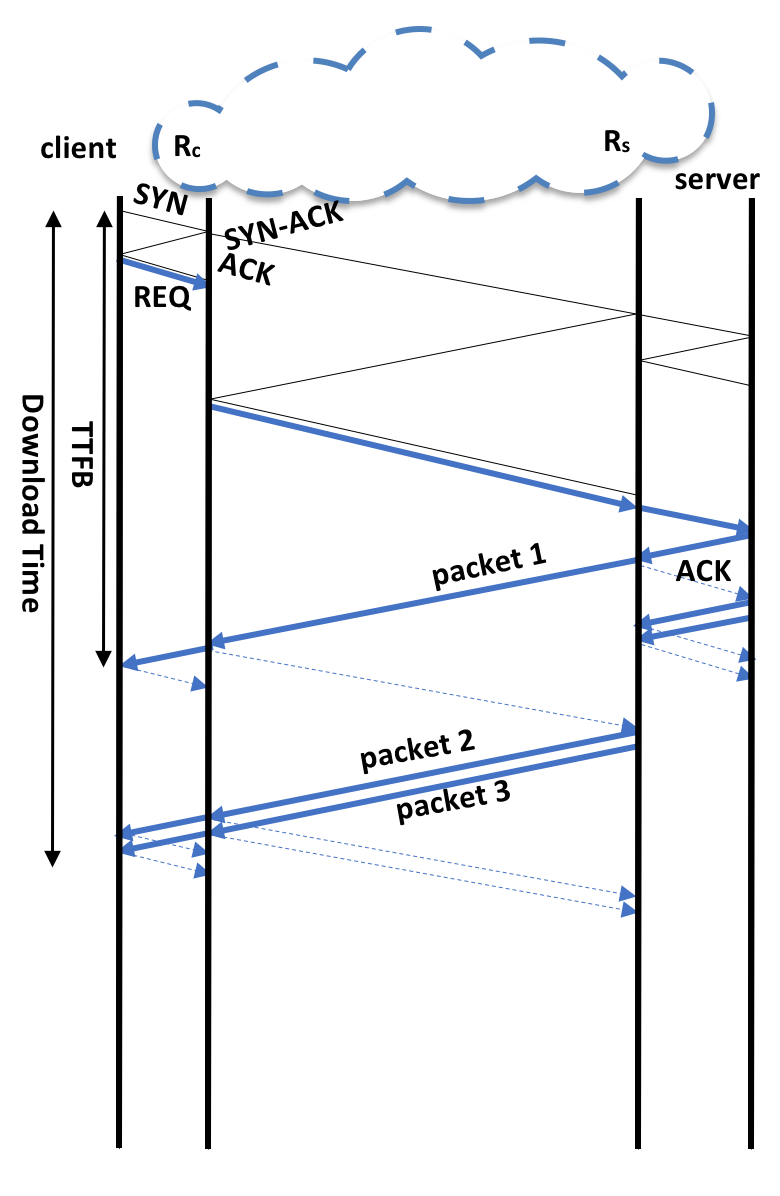}
    \caption{Thread pool.} \label{fig:thread-pool}
\end{subfigure}    \centering
\begin{subfigure}{0.48\columnwidth}
  \centering
  \includegraphics[width=\columnwidth]{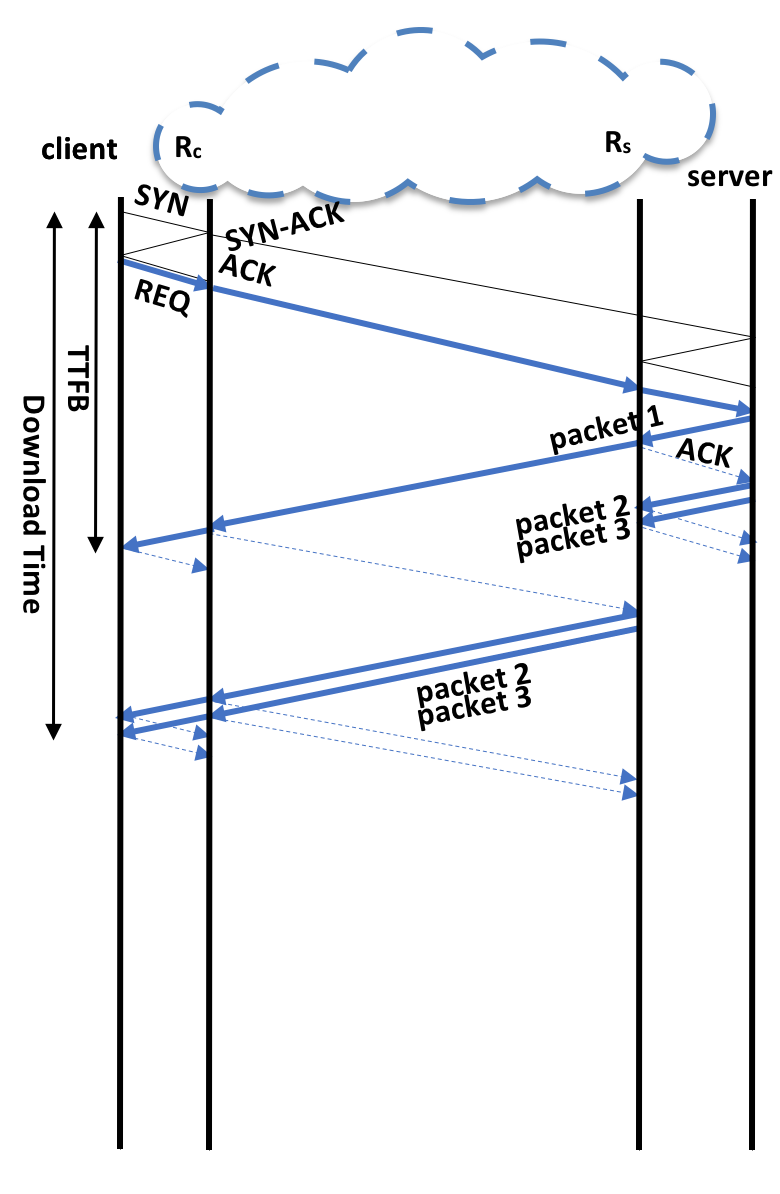}
    \caption{Connection pool.} \label{fig:connection-pool}
\end{subfigure}     \centering
\begin{subfigure}{0.48\columnwidth}
  \centering
  \includegraphics[width=\columnwidth]{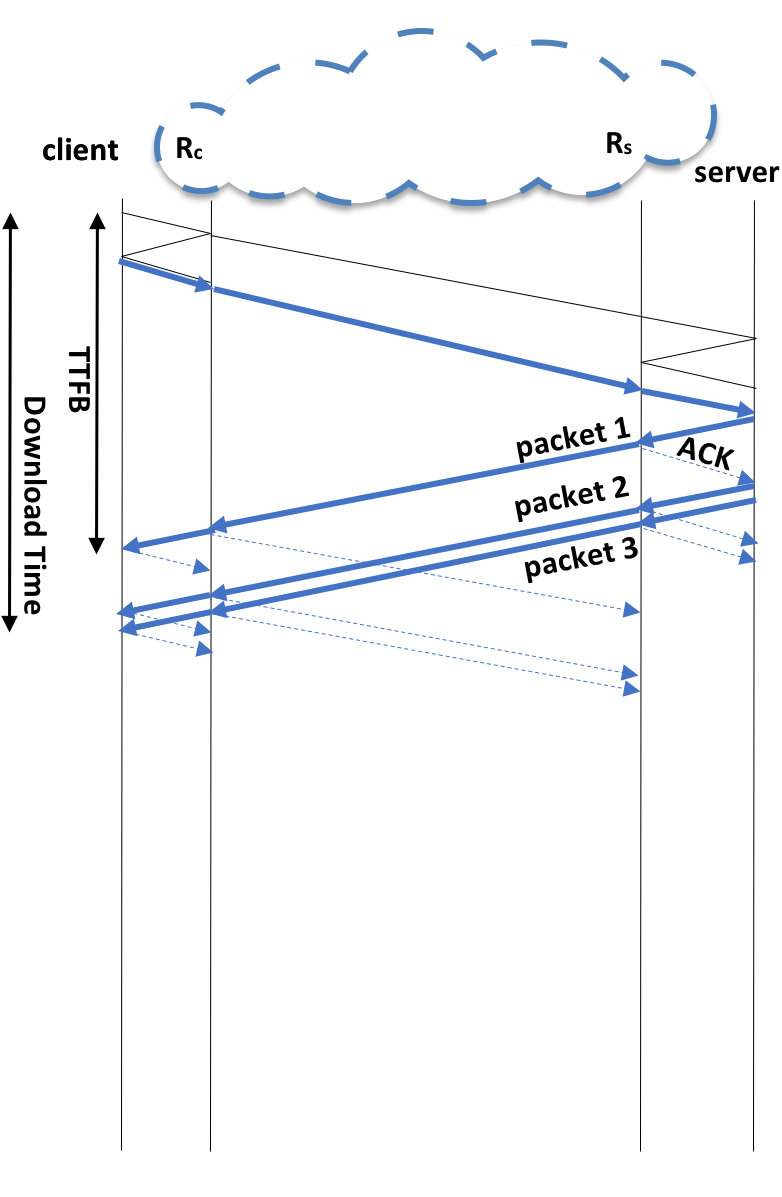}
    \caption{Turbo-Start TCP.} \label{fig:turbo-start-tcp}
\end{subfigure}
    \caption{\oursys successive implementation improvements: Using \textit{Early-SYN}, we can remove SYN-ACK and ACK delays (marked as (1) in Figure~\ref{fig:baseline}). Using a \textit{thread pool} removes forks (marked as (2) in Figure~\ref{fig:baseline}). With a \textit{connection pool}, delay (3) in Figure~\ref{fig:baseline} is eliminated. Turbo-Start TCP eliminates delay (4). The two last delays from Figure~\ref{fig:baseline} that need be removed are delays $\Delta_c$ and $\Delta_s$ on the client and server sides, respectively. As both depend on the client and server parameters, they seem beyond our control. 
    }
    \label{fig:KSPLIT-improvements}
\end{figure*}

\T{Congestionless control?} The cloud provides an auto-scaling networking infrastructure with links of virtually-infinite capacity. Like others~\cite{haq2017measuring}, we find that in-cloud paths provide more predictable results than the public Internet, with an order of magnitude lower loss rate. As a result, flows in the cloud will rarely ever encounter congestion. This compels us to rethink the role of congestion control in the cloud. In light of the near-absence of congestion, we essentially want a simple high-rate sending protocol capable of dealing with the very rare loss events, without necessarily backing off. Even flow control is not necessarily needed within the cloud, as clouds can increasingly provide an elastic memory infrastructure~\cite{hotadd,baloon} that can quickly adapt to any receive-buffer temporary load. 

\T{Ideal pipe.} Given the large capacity of the cloud, we wonder how close we could get to an ideal transmission if we were to use a cloudified data transfer.
\autoref{fig:ideal} illustrates our fundamental model of an ideal transmission. Importantly, this model reflects a protocol-free, theoretical ideal transmission. Our aim for the remainder of this section is to identify the means for approximating this model in practice. We show in \autoref{fig:e2e} that a classical end-to-end data transfer based on HTTP over TCP falls short 
of this goal. In addition, the OCD Baseline strategy of Section~\ref{sec:ocd-baseline} also introduces additional delays with respect to the ideal model. We point out these delays in \autoref{fig:baseline}, and discuss how these delays can be eliminated, so as to come close to our theoretical ideal.

\subsection{Approximating the Ideal Pipe}\label{sec:approx}

To approximate the ideal data transmission model of Section~\ref{subsec:dream-pipe}, we introduce \textit{\oursys}.
The goal of \oursys is to enhance the OCD Baseline of Section~\ref{sec:ocd-baseline} and provide efficient, delay-free TCP optimization; while utilizing commodity VMs and standard programming APIs. This is done by incorporating four improvements to the baseline strategy, illustrated in Figure~\ref{fig:KSPLIT-improvements}. Together, these four components eliminate the delays marked (1)-(4) in Figure~\ref{fig:baseline}. We next elaborate on each of the four improvements. We discuss the many implementation details involved in realizing \oursys in Section~\ref{sec:design} and provide our open-source code in~\cite{ktcp}.

\T{Improvement 1: Early SYN.} In early SYN~\cite{ladiwala,siracusano2016miniproxy}, a SYN packet is sent to the next-hop server as soon as the SYN packet arrives. This is done without waiting for the three-way handshake to complete. \oursys captures this first SYN packet and triggers the start of a new connection. This allows the proxy to establish the two legs of a split connection in parallel. (Note that while this improvement is well-known, to our knowledge the next ones are novel in the context of TCP switching.)

\T{Improvement 2: Thread pool.} 
The creation of new kernel\_threads for each new split connection is time-consuming and adds greatly to the connection jitter. Some outliers may take tens of milliseconds, greatly hurting performance. For small files/objects used by a time-sensitive application, this jitter may even nullify the benefit of \oursys. To mitigate this problem, we create a pool of reusable kernel threads. These are sleeping threads, awaiting to be assigned to new tasks.

\T{Improvement 3: Reusable connections.}  This optimization aims to improve the performance over long connections, \ie those where the RTT between the two cloud relays dominates. The goal is to negate the delay of the long three-way handshake. We achieve this goal by preemptively connecting to the distant \proxies. In fact, we pre-establish a pool of \reconn between each pair of distant \proxies (the cost is quadratic in the number of \proxies).

\T{Improvement 4: Turbo-Start TCP.} 
Congestion is not an issue within the cloud, hence, there is essentially no need to use TCP's slow-start mechanism. It is redundant to probe the network when a connection is established between two \proxies within the same cloud provider. We thus configure a large initial congestion window (CWND) and large receive window (RWIN) on \oursys \proxies. In addition, we increase the socket buffers for the relay machines, so that memory would not limit the performance of the intra-cloud flows.
Note that we do not change the CWND used on any Internet-facing flows. We wish to remain friendly to other TCP flows potentially sharing a bottleneck link with our \proxies. 

\subsection{Experimental Evaluation}\label{subsec:improving-baseline}
To evaluate the contribution of each of these improvements, we set up a server in Bangalore as a VM on a Digital Ocean infrastructure; and a client PC in San-Francisco, connected using a residential Internet service \footnote{The ISP is Comcast.}. Our relays are two VMs on GCP's public cloud: one in Mumbai, close to the server (\rs), and another (\rc)  in Oregon, near the client. Both VMs run on Ubuntu 17.04 and use small (n1-standard-1) machines with a single vCPU and $3.75$ GB memory. The client and server are Ubuntu 16.04 machines.

\begin{figure}
  \centering
    \includegraphics[width=\columnwidth,trim=20mm 25mm 20mm 22mm,clip]{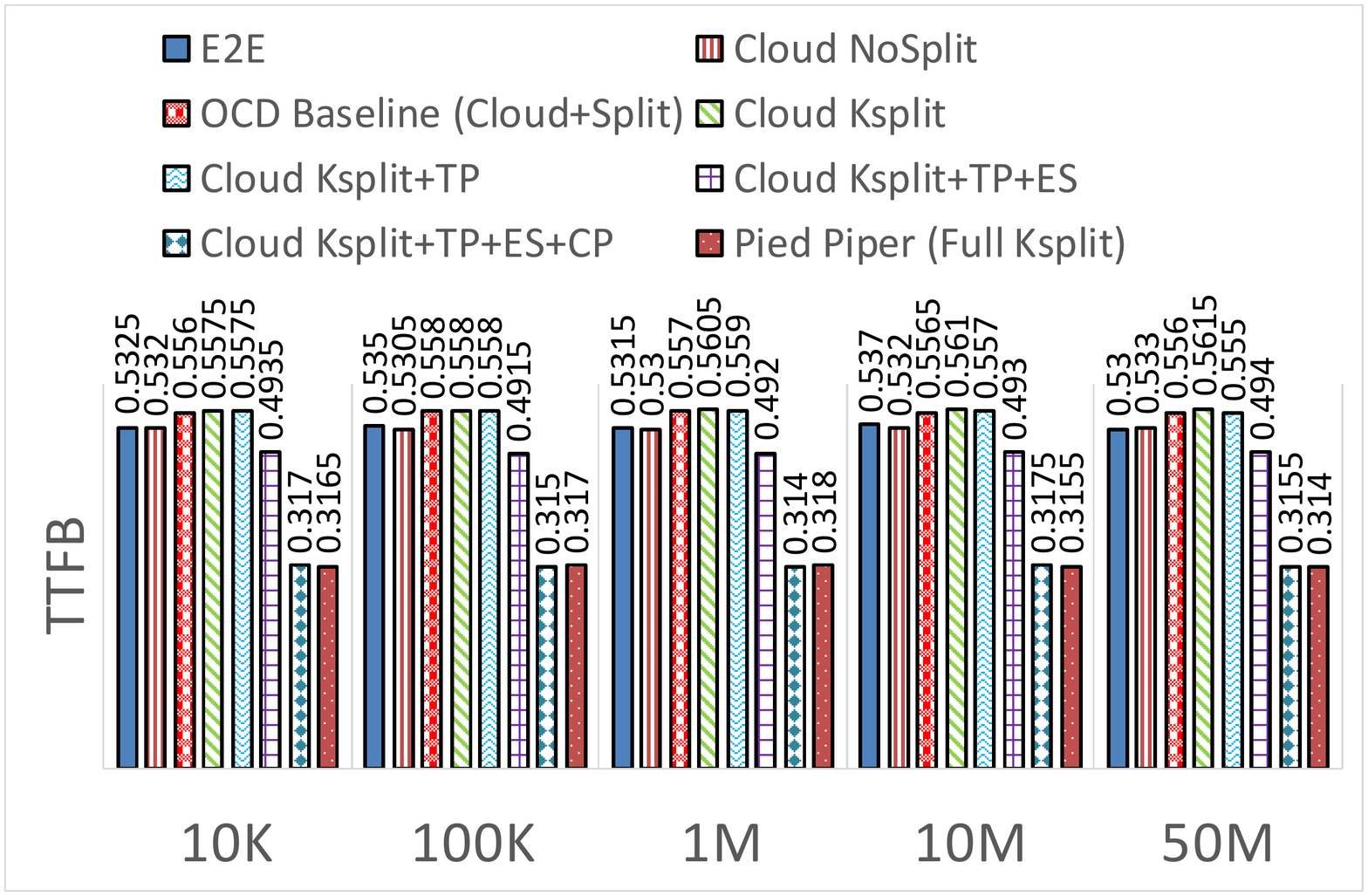}
    \caption{Time-To-First-Byte (TTFB) in seconds. Median results of 50 runs.}
    \label{fig:ksplit-ttfb}
\end{figure}

\begin{figure}
  \centering
    \includegraphics[width=\columnwidth,trim=20mm 25mm 20mm 23mm,clip]{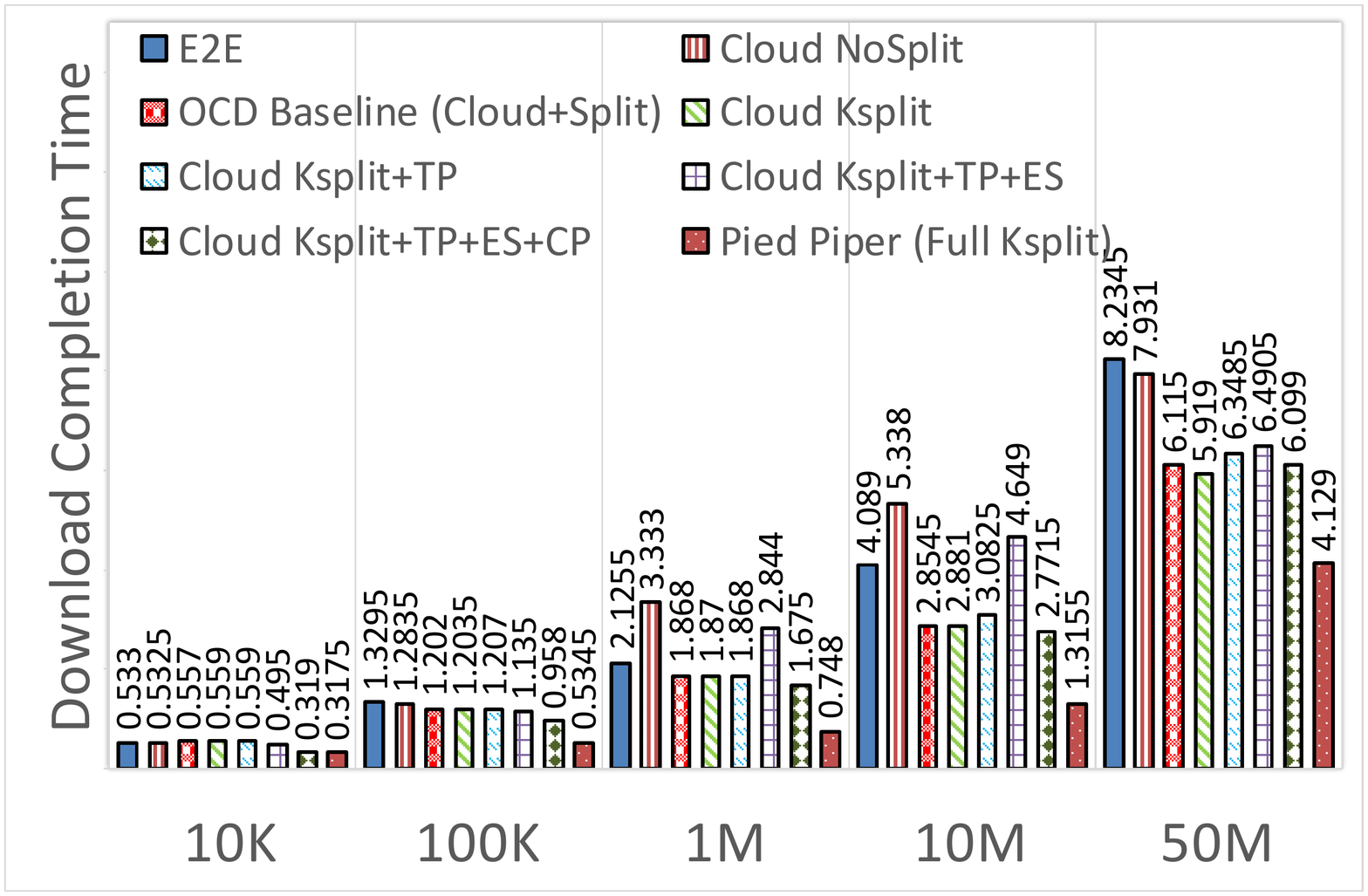}
    \caption{Download completion time in seconds. Median results of 50 runs.} 
    \label{fig:ksplit-download}
\end{figure}

We set up an Apache web server on the Bangalore VM and evaluate the performance of each of the options by measuring both download times and time-to-first-byte (TTBF). We experiment with files of different sizes that the client is downloading from the server, using HTTP via the curl utility. The times reported in \autoref{fig:ksplit-ttfb} and \autoref{fig:ksplit-download}  are as returned by the curl utility. 

The RTT  (as measured by ICMP) between the client and $\rc$ is $32.7$ms, between $\rc$ and $\rs$ is $215$ms, and between $\rs$ and the server is $26$ms.

We compare the performance of the following configurations: \begin{romanlist}
     \item simple End-to-End (\textit{e2e});
     \item routing through the cloud relays; using iptable's DNAT, without splitting the TCP flow (Cloud NoSplit);
     \item splitting the TCP flow using SSH's port forwarding feature (\textit{OCD Baseline});
     \item TCP splitting using our \ksplit kernel module, set up to use the improvements listed in \autoref{sec:approx}: thread pool only (Cloud \ksplit+TP), thread pool and early-SYN (Cloud \ksplit+TP+ES), a complete bundle of the three connection improvements including \reconn (Cloud \ksplit+TP+ES+CP), and finally also configuring the intra-cloud TCP connection to use Turbo-Start (\textit{\name}).
\end{romanlist}

The benefit from the improvements is best observed by looking at the Time-To-First-Byte in \autoref{fig:ksplit-ttfb}. We can see that the  TTFB and total download time of the basic \ksplit coincide with those of the OCD Baseline. Our basic kernel-based implementation performs at least as well as the well-established ssh utility. We also note that \ksplit+TP does not improve the median performance by a noticeable amount. However, we have noticed throughout our testing that the thread pool improves the stability of our results.

For all file sizes we notice an improvement of $\approx60ms$ when using \ksplit+TP+ES. This is in line with \autoref{fig:baseline}, as Early-SYN eliminates one RTT on the $(\mbox{client}\leftrightarrow\rc)$ and another RTT of the $(\rs\leftrightarrow\mbox{server})$. The amount of time reduced, according to our RTT measurements is supposed to be $59$ms, in line with our results. Adding \reconn to the mix should potentially reduce the TTFB by one RTT of the $(\rc\leftrightarrow\rs)$ leg. However, since the REQ cannot be forwarded with the SYN packet sent by the client (without any TCP extensions), we can only gain $215-33=182$ms. Indeed, the benefit of adding CP as evident in \autoref{fig:ksplit-ttfb} is of $\approx 180$ms. The addition of Turbo-Start does not reduce the TTFB, as it only influences the way packets are sent after the first bytes. The contribution of Turbo-Start is clearly evident when considering the total download time (\autoref{fig:ksplit-download}). We see considerable reduction of the file download time when using Turbo-Start for all file sizes, except that of 10 KB file. The default initial congestion window size for both Ubuntu 17.04 and 16.04 is 10 segments, so the entire file is sent in a single burst. Indeed, the results show that for 10 KB file the download completion time is about 1 ms after the first byte arrives. All other improvements contribute to the reduction of TTFB, and so reduce the total download time by roughly the same amount. This reduction is barely noticeable for large files, where the main benefits stem from splitting the TCP flow and using Turbo-Start.
In this experiment we notice that the best performing implementation improvement (\ie \oursys) outperforms e2e file transfer by up to 3 times! (depending on the file size).

\begin{figure}
  \centering
    \includegraphics[width=\columnwidth,trim=20mm 25mm 25mm 25mm,clip]{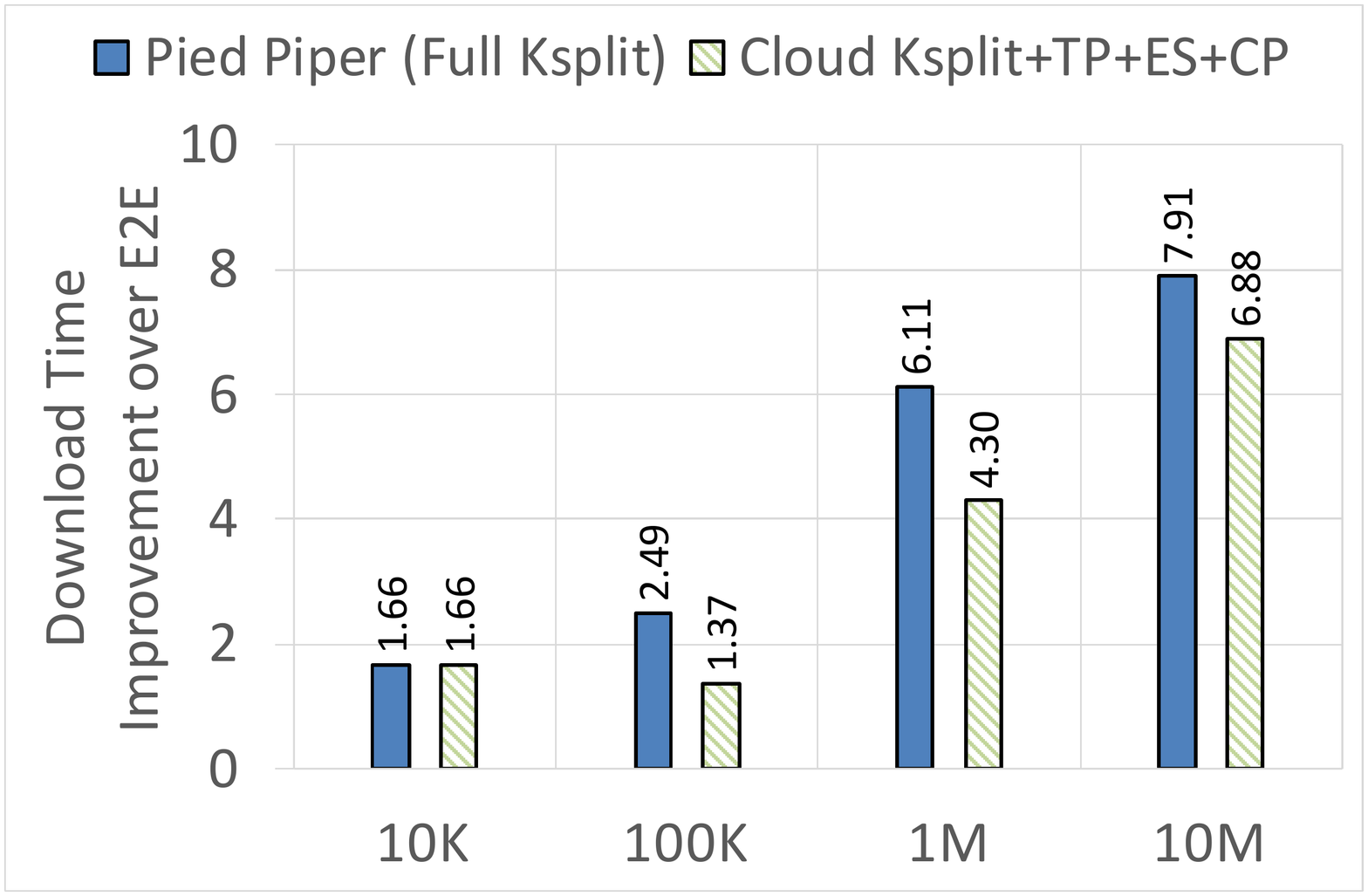}
    \caption{Improvement of \ksplit performance over e2e in the case of a memory-limited client. Averaged results over 30 runs.} 
    \label{fig:ksplit-weak-client}
\end{figure}

We also consider what benefit OCD can present to clients with limited memory, such as old PCs, small battery- and/or price-limited devices as well as other clients on the Internet which might advertise a limited receive window. Google noted in their QUIC paper \cite{quic} that 4.6\% of the clients downloading video from their servers (March 2016) had a limited receive window of 64KB. For large-RTT flows, this limited receive window directly limits the maximum throughput of a TCP flow. This is due to TCP's flow-control mechanism. OCD reduces the RTT of the first hop considerably,\footnote{compared to the original e2e RTT.} and thus potentially reaches a higher throughput with the same small receive window.
To assess the benefits such clients might gain from using OCD, we rerun \oursys and K-Split+TP+ES+CP experiments and compare their performance when the client's receive socket buffer is limited to 64 KB. The results are presented in \autoref{fig:ksplit-weak-client}. We see up to a $7.91\times$ speed up when using \oursys for clients with a limited receive buffer. For 10 KB files the improvement is much more modest, as the receive window is not the limiting factor. In this case, all the benefits are due to the TTFB reduction. 

To summarize, we demonstrated in this section how we can, with some acceleration techniques within our kernel implementation of TCP split, improve download times of files  of all sizes, and also improve latencies by reducing TTFB.

\section{OCD Evaluations
}\label{sec:baseline-good_enough}

\subsection{Two Relays are Enough}\label{subsec:num_of_relays}
When using an overlay network, one may ask, how many relays do we need? In the following evaluations, we began by trying all the options for a single relay. We compared this to our Baseline approach with two relays, in which one is geographically-closest to the client, and the other geographically-closest the the server. Finally, we  tested adding the nodes that performed best as single relays to these two relays.

\begin{figure}
  \centering
    \includegraphics[width=\columnwidth,trim=2mm 5mm 1mm 5mm,clip]{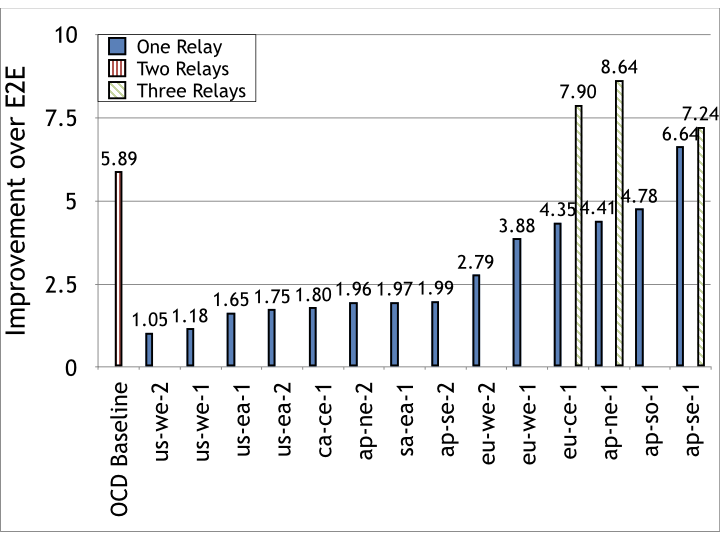}
    \caption{One, two and three relays when using TCP splitting in AWS for the Mumbai-SF international setting. (1)~The first striped red bar shows the performance of the simple OCD Baseline two-relay placement at the nodes that are closest to the client and the server. (2)~Next, the 14 blue bars show all the options for a single-relay option. The best option is ap-se-1 with a normalized performance of 6.64$\times$, above the Baseline two-relay option. However, all other single-relay options have a lower performance, and the median option gets 1.99$\times$, \ie only 30\% of the best performance. (3)~Finally, the three green bars show a few options for the middle relay in the three-relay method, given the Baseline two-relay options at the extremities; \eg eu-ce-1 shows the performance of consecutively going through (us-we-1$\to$eu-ce-1$\to$ap-so-1). The best option is ap-ne-1 with an 8.64$\times$ performance. } 
    \label{fig:how-many-hops-aws}
\end{figure}

\autoref{fig:how-many-hops-aws} extends a portion of \autoref{fig:must-split}, showing the results for all the relay options tested in AWS using TCP splitting in the India - West Coast route. 
We learn that (1) all but one relay perform \emph{worse} than the simple two-relay option; (2) the \emph{best} single relay (ap-se-1, \ie ap-southeast-1 in Singapore) indeed performs better than our simple two-relay approach, but this relay is neither the closest to the client, nor the closest to the server, nor the (geographically) closest to the mid-point. 
This suggests that determining the identity of the best single relay to use is nontrivial as this relay seems to have no obvious characteristics. 
(3) Using three  relays can typically (for the proper choice of relays) strictly improve over the 2-relay strategy, but it is not obvious to select the correct three relays in advance.

While \autoref{fig:how-many-hops-aws} shows results for relays in AWS, similar results were observed when using GCP and Azure.

\subsection{TCP is Enough} 
While the effects of congestion control in overlay networks have been discussed before \cite{CRONets}, much is left unknown.
We tested the performance of the default TCP Cubic against BBR and Turbo-Start Cubic with the same international setting as in \autoref{fig:how-many-hops-aws} on AWS using TCP splitting. Note that these results reflect the deployment of BBR and Turbo-Start Cubic \textit{only on the relay nodes} and not at the end-points.

\begin{figure}
  \centering
    \includegraphics[width=0.47\textwidth,trim=2mm 2mm 2mm 2mm,clip]{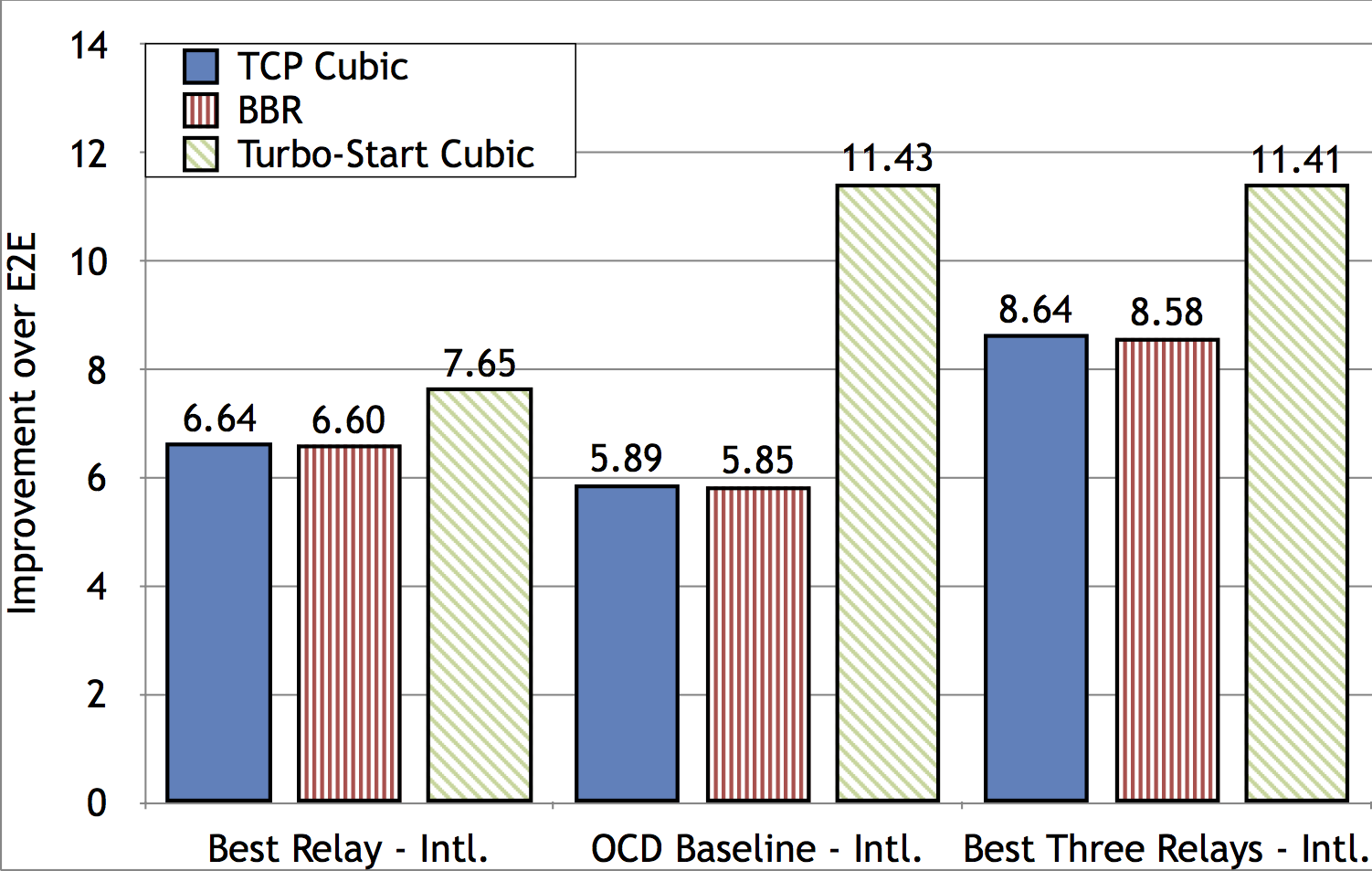}
    \caption{
    The impact of using different congestion control options is demonstrated using the Mumbai-SF international transmission of a large file with TCP splitting over AWS. In all three cases of best-single, OCD Baseline and best-triple relays, Turbo-Start Cubic achieves remarkably higher efficiency. In the last two cases, it achieves an impressive 11.4$\times$ improvement over e2e.
   }
    \label{fig:aggressive}
\end{figure}

\autoref{fig:aggressive}
showcases representative results. Here we utilize the residential client in the West Coast, relays in AWS, and a web server in Mumbai. As can be seen, BBR did not achieve better performance than default TCP.

\subsection{Using Turbo-Start TCP Pays Off}\label{subsec:quick-start}
\autoref{fig:aggressive} illustrates the impact of using Turbo-Start TCP Cubic  instead of TCP Cubic or BBR.
We observe that (1) Turbo-Start TCP significantly improves upon TCP Cubic and BBR, and (2) this improvement is particularly large with two or three relays. We hypothesize that their performance gain is similar because the main benefit of using the third (middle) relay is in reducing the impact of the slow-start phase by reducing the RTT, and Turbo-Start TCP avoids slow-start altogether.

\subsection{Is RTT a Good Proxy for Performance?}

In contrast to measuring bandwidth, measuring latency (in terms of RTT) is simple. When is RTT a good proxy for performance, in terms of download completion time?

\begin{figure}[t]
  \centering
  
    \begin{subfigure}{0.47\columnwidth}
  \centering
  \includegraphics[width=\columnwidth]{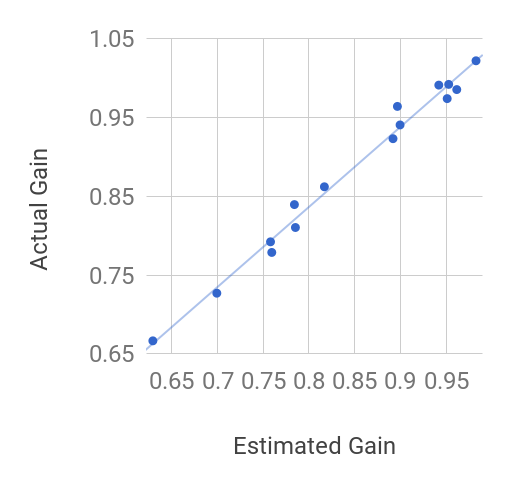}
    \caption{}
    \label{fig:rtt-estimate-nat-1hop}
\end{subfigure} \hfill
\begin{subfigure}{0.47\columnwidth}
  \centering
  \includegraphics[width=\columnwidth]{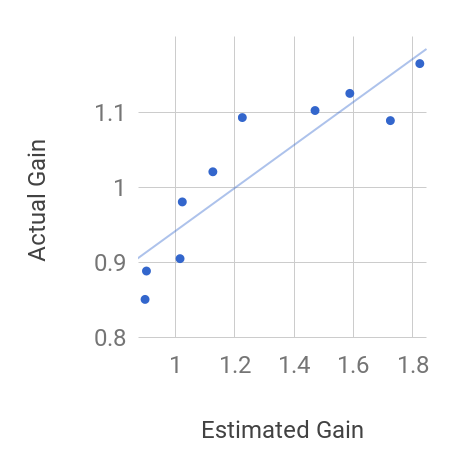}
    \caption{} \label{fig:rtt-estimate-ssh-3hop}
\end{subfigure}
    \caption{Performance gain estimates \vs actual gain of \textbf{(a)} the single relay method without splitting; and \textbf{(b)} the 3 relay setup with splitting.}
\end{figure}

\noindent{\bf Without TCP splitting.} When TCP is \emph{not} split, end-to-end RTT is a good proxy for download completion time: the shorter the RTT, the shorter the download. We used RTT measurements to estimate the improvement of different routing strategies over e2e by calculating the ratio between the e2e RTT and the RTT of the tested route. \autoref{fig:rtt-estimate-nat-1hop} plots the correlation between our estimator and the actual performance gain. 

\noindent{\bf With TCP splitting.} When TCP splitting is used, with Cubic as the congestion control algorithm, an RTT-measurements-based estimator succeeded in predicting the performance gain of the 3-relay method over the 2-relay method. This is accomplished by calculating the ratio of the RTT between \rc and \rs and the maximum RTT of the segments of the route through the third node used between them. Namely, if $\rtt(\rc,\rs)$ is the RTT between \rc and \rs, $\rtt(\rc,\rmid)$ is the RTT between \rc and the middle relay, and $\rtt(\rmid,\rs)$ is the RTT between \rs and the middle relay, we estimate that using this third relay would improve download time by 
\[
\frac{\rtt(\rc,\rs)}{\max(\rtt(\rc,\rmid),\rtt(\rmid,\rs))}
\]
compared to using the two-relay route using \rc and \rs only.
The reasoning behind this estimator is that when splitting is used, the throughput is limited by the slowest segment of the route. \autoref{fig:rtt-estimate-ssh-3hop} depicts the correlation between this estimator and the actual gain of the 3-relay method over the 2-relay method (both with TCP splitting at each relay).

Motivated by the relative accuracy of this estimator, we used RTT measurements to predict which 4-relay route would fare best when using TCP splitting and Cubic as the congestion control protocol. The 4-relay route achieved an improvement of 8.57$\times$, \ie higher than the 7.38$\times$ improvement achieved with the 3-relay route calculated with min-max RTT between \rc and \rs. 
 
We point out that, in contrast to the above results, this estimator fails to provide good results for the single relay with TCP splitting strategy, possibly because as RTTs get higher the impact of congestion and loss rate becomes more significant.

\subsection{Mixing Clouds}\label{subsec:mixing-clouds}

We also tested whether routes traversing different clouds can produce better results. To this end we rerun the 2-relay and 3-relay experiments (with TCP splitting) using the SF client and Mumbai server with \rc on AWS and \rs on Azure. We compared performance to that of the same strategies with AWS relays only. Our preliminary results show that mixing clouds was indeed beneficial in some of the evaluated scenarios--- specifically, when Turbo-Start Cubic was used and/or when three relays were used (with either Cubic or Turbo-Start Cubic).

\section{OCD \ksplit Implementation}\label{sec:implementation}

\label{sec:design}
\T{\ksplit.} We have developed a novel kernel module called \textit{\ksplit} in Ubuntu 17.04, and have made it openly available on Github~\cite{ktcp}. \ksplit implements a kernel-based TCP split together with the four improvements of OCD \name over the OCD Baseline (\S\ref{sec:rate-control}). 
In addition, \ksplit enables utilizing easily-deployable commodity VMs, as well as  standard programming APIs (POSIX/\sockets). 

\T{Kernel mode.} We implemented \ksplit in kernel mode. We rely on procfs~\cite{proc} to control the behaviour of \ksplit. A virtual file system~\cite{virtfs} provides a simple interface and facilitates easy scripting; additionally, it allows to communicate in run time with \ksplit.

The decision to use kernel mode is a significant one. While developing in user space would have provided an easier development environment, implementing \ksplit in the kernel allows us to (1) take advantage of resources only available in the kernel, such as \textit{Netfilter}~\cite{netfilter}, which is crucial to our needs; and (2) avoid the penalties that stem from numerous \textit{system calls}~\cite{Copy, FlexSC}. By working in the kernel, we eliminate the redundant transitions to and from user space and avoid gratuitous system calls.

The decision to implement the components of \ksplit in the kernel is further made easy by the fact that all socket APIs have kernel counterparts. One limitation of an in-kernel implementation is that epoll~\cite{epoll}, a scalable I/O event notification mechanism, has no kernel API. Instead, we use kernel\_threads to service our sockets. 
We have measured the cost of context switching between kernel\_threads on our setup, assuming an n1-standard(GCP) VM with Intel Skylake Xeon CPU. We measured the time it takes two kernel\_threads to call schedule() 10 million times each. This experiment completes in under 3.2 seconds, resulting in 0.16 \usec per context switch on average; by comparison, an analogous experiment with two processes runs for 15.6 seconds and 12.9 seconds for two POSIX threads~\cite{pthreads}.
Of course, these are not exactly comparable because the user-space experiments invoke system calls in order to switch context (sched\_yield()). In any case, since we want to avoid system calls and minimize context switches, implementing \ksplit in the kernel is the logical choice.

\T{Basic implementation.} The basic implementation of \ksplit relies on three components. (1) We create a socket that listens for incoming connections. (2) Iptable~\cite{iptables} rules redirect specific TCP packets to our proxy socket. (3) A second TCP socket is used to connect to the destination and thus complete the second leg of the split connection. Once both connections are established, the bytes of a single stream are read from one socket, and then forwarded to its peer. This forwarding happens in both directions. When either connection is terminated via an error or FIN, the other connection is shut down as well. This means that the bytes in flight (\ie not yet acked) will reach their destination, but no new bytes can be sent.

\T{Buffer size.} We found that the size of the buffer used to read and write the data is important. At first we used  a 4KB buffer, and experienced degraded performance. However, 16KB and 64KB maintain the same stable speed.

\T{Implementing Early-SYN.} 
As there is no standard API that enables the capture of the first SYN packet, we use Linux Netfilter~\cite{netfilter} hooks. We add a hook that captures TCP packets, and then parse the headers for the destination and the SYN flag. With this information \ksplit launches a new kernel\_thread\footnote{Creating a new thread while in the Netfilter callback is not allowed. We use our thread pool to lunch kernel\_threads from atomic contexts.} that initiates a connection to the intended destination. Capturing the SYN allows the \proxies to establish the two sides of a connection concurrently.

\T{Implementing the thread pool.} 
Each split connection is handled by two dedicated kernel\_threads~\cite{kthread}. Each thread receives from one socket and writes to its peer. One thread is responsible for one direction of the connection. We use blocking send/receive calls with our sockets allowing for a simple implementation; this also means that we need a kernel\_thread per active socket. Unfortunately, the creation of a new kernel\_thread is costly. On our setup, a kernel\_thread creation takes about 12\usec, on average.\footnote{By comparison a fork consumes more than 25\usec , while launching a POSIX pthread consumes around 13\usec.} But an outlier may consume several milliseconds, resulting in a jittery behaviour.

To mitigate this problem and the problem of creating new kernel\_threads from atomic context, we create a pool of reusable threads. Each kernel\_thread in this pool is initially waiting in state TASK\_INTERRUPTIBLE (ready to execute). When the thread is allocated, two things happen: (1) a function to execute is set and (2)  the task is scheduled to run (TASK\_RUNNING). When the function is finished executing, the thread returns to state TASK\_INTERRUPTIBLE and back to the list of pending threads, awaiting to be allocated once more. A pool of pre-allocated kernel threads thus removes the overhead of new kernel\_thread creation. A new kernel\_thread from the waiting pool can start executing immediately and can be launched from any context. When the pool is exhausted, \ie all pool threads are running, a new thread will be allocated; thus for best performance the pool-size should be configured to cover the maximum possible number of concurrent connections. On a multi-core system, the heavy lifting of thread creation is offloaded to a dedicated core. This core executes a thread that allocates new kernel\_threads any time the pool size dips under some configurable value. On the other hand, when threads return to the pool, it is possible to conserve system resources by restricting the number of threads awaiting in the pool and freeing the execs threads.\footnote{Each kernel\_thread consumes 9KB of memory.}

\T{Implementing the \reconn.} 
To implement \reconn, we have added a second server socket. Unlike the ``proxy'' socket, this socket listens for connections from other \proxies that are initiating new \reconn. In order to keep the connection from closing before it is allocated, the sockets are configured with KEEP\_ALIVE.

When established, these connections await for the destination address to be sent from the initiating peer. The destination address is sent over the connection itself. This information is sent in the very first bytes, and all following bytes belong to the forwarded stream. 
Once the destination address is received, a connection to the destination is established and the second leg of the split connection is established. The streams are forwarded between the sockets just like in the basic design.

We found that Nagle's Algorithm~\cite{nagle} should be disabled on these sockets. In our experiments, we have seen that without disabling it, the time-to-first-byte is increased by some $200$ milliseconds.  

\T{Proc.} The size of the thread-pool, the destination of a \reconn, and their number are controlled via the procfs~\cite{proc} interface.

\T{Effort.} The total implementation of \ksplit is less than 2000 LOC (lines of code). The basic implementation is about 500 LOC, thread pool and early syn add 300 LOC each, and \reconn add 500 LOC. The code for the proc interface and common utilities consists of about 300 LOC.

\T{Implementation at scale.} We now want to briefly discuss how the existing implementation may be scaled in the future. With 10K split connections, the memory footprint of socket buffers alone; far exceeds the size of the shared L3 cache of most modern servers\footnote{On GCP, it is an impressive 56MB.}. It may be prudent to expand the epoll API to the kernel and thus save the 18KB of memory per split connection. Extending epoll will not be enough, other avenues should be considered as well.
One such idea is the socket API; socket API in the kernel is not zero copy. The needless copy can become detrimental ~\cite{Copy} due to an increase in costly memory traffic. Another point to consider is that, network I/O is serviced by interrupts. For a virtual machine, this means expensive VM exits~\cite{Eli, Elvis}. It is well documented that para-virtual devices like~\cite{virtio,vmxnet3} have sub-optimal performance~\cite{Eli, Elvis}. An SRIOV device and a machine with a CPU that supports Intel's vt-d posted interrupts~\cite{posted} may be needed to achieve near bare-metal performance.

\section{Deployment Strategies}\label{sec:deployment}

We next discuss strategies for OCD deployment (somewhat similar to the suggestions in \cite{pucha2005slot}). We consider two different contexts: (1) a cloud provider deciding to provide OCD as a service, and (2) another entity (say, a website) deciding to employ OCD for its traffic. 

\subsection{Enhancing Cloud Services with OCD}\label{subsec:enhancing-w-ocd}

Similarly to the transition to cloud computing, cloud service providers such as Google, Amazon, and Microsoft, can leverage their resources (Data Centers, network capacity) to broaden their business offerings, by providing transit services.

Realizing OCD by a cloud provider can be accomplished by utilizing mechanisms employed by CDNs, such as the following. Suppose that a content owner such as CNN wishes traffic to/from the content to be sent via OCD, and purchases OCD services from the cloud. The cloud service provider can then set up cloud relays in the vicinity of the CNN content (and autoscale them as needed). Now, consider a client $A$ located outside the cloud. When the client seeks CNN's IP address through DNS, the corresponding DNS records should be overridden so as to point that client to the IP address of the cloud relay closest to that client. Then, traffic from the client to CNN and vice versa can be sent through the cloud by entering/leaving the cloud via the corresponding cloud relays.

\subsection{Incremental Deployment of OCD}

Now, consider the scenario of a content owner such as CNN that wishes to use OCD. Yet, no cloud service provider offers OCD as a service. CNN, in this example, can deploy OCD for its own traffic using the exact same methods as described in Section~\ref{subsec:enhancing-w-ocd}. Of course, the overhead of doing so is non-negligible (overriding DNS records, setting up virtual machines in the cloud and auto-scaling them, \etc). Alternatively, non-cloud-service-providers can offer such services to content owners (such as CNN). These non-cloud-service-providers potentially could create a virtual overlay over the cloud and solve the associated operational challenges.

\subsection{Fairness}

\begin{figure}[!t]
  \centering
    \includegraphics[width=0.46\textwidth,trim=20mm 25mm 25mm 20mm,clip]{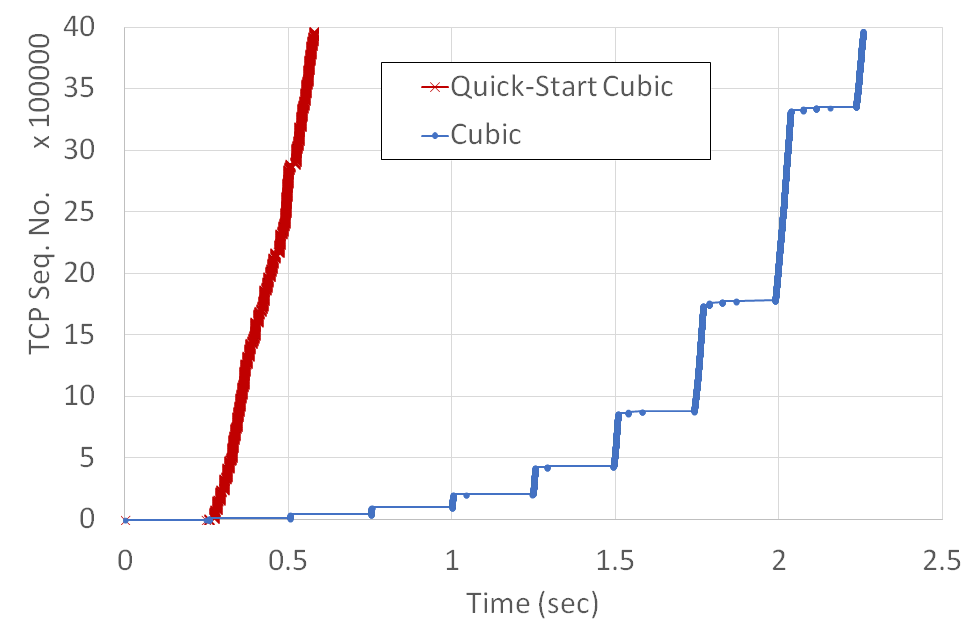}
    \caption{Per-byte look at the flow between \rs and \rc, as captured on \rc. The graphs depict the sequence number received \vs relative time (to the beginning of the TCP connection). The transmission rate during the bursts in the default cubic option is 260 MBps while it is only 110 Mbps when the Turbo-Start option is used. However, since Turbo-Start avoids slow-start, its overall download time is much shorter.  
    }
    \label{fig:agg-vs-cubic}
\end{figure}

\T{Inside the cloud.} We were wondering whether Turbo-Start Cubic is too aggressive and unfair towards other regular TCP flows by essentially skipping the slow-start phase of TCP. Even though we keep the connection remainder unchanged, aren't we just hogging the link and exploiting a loophole
that cloud providers should soon close?
\autoref{fig:agg-vs-cubic} looks at the per-byte download time of our large file in the cloud. We learn that (1) the average throughput is clearly higher for the Turbo-Start Cubic, as expected from its better performance (\autoref{fig:aggressive}); but (2) this is mainly because TCP Cubic tends to have abrupt packet bursts during its slow-start phase. First waiting a full RTT for a batch of ACKs to arrive back to the server, then sending bursts of packets to catch up. When zooming on the TCP-Cubic bursts, it is clear that the gradient is steeper than that of Turbo-Start Cubic, \ie it is \textit{more} aggressive because of this tendency to catch up. 

\T{Outside the cloud.} We also considered the legs of the route that reside outside the cloud, \eg between the client and the relay \rc closest to it. Are we somehow hogging the Internet resources to achieve better performance? 

Our design focuses on improvements within the cloud relays, and only modifies TCP parameters for the intra-cloud legs of the route. We deliberately make sure the Internet-facing flows originating from OCD relays use the default Linux settings. Any TCP traffic from a \oursys relay should compete fairly with other TCP flows, just  like any other flow on the Internet.

\begin{figure}[!t]
  \centering
    \includegraphics[width=0.46\textwidth,trim=20mm 25mm 25mm 20mm,clip]{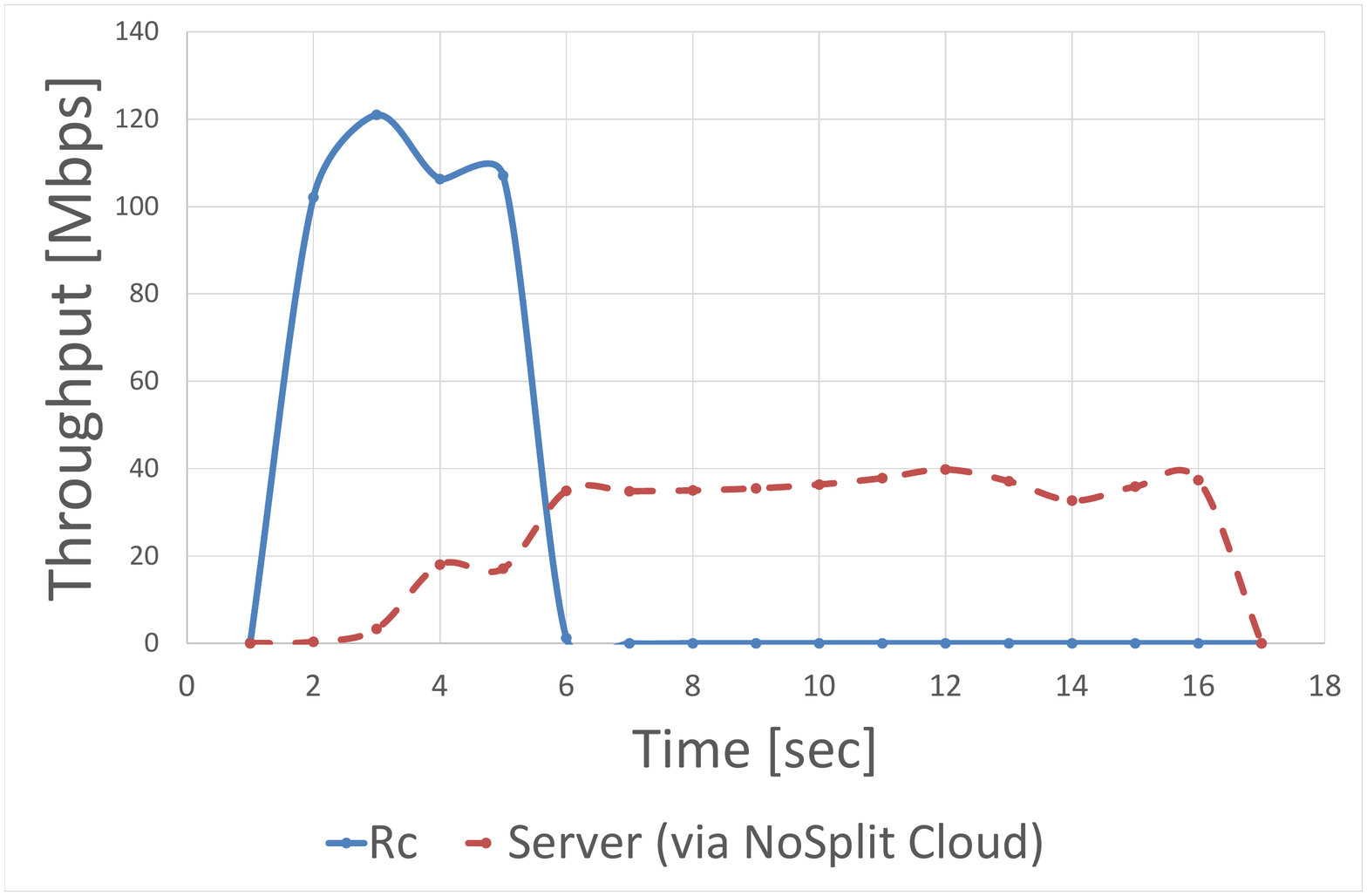}
    \caption{ (Un)Fairness between two competing flows destined to the same client in San Francisco. One is downloading a 50 MB file from $\rc$, with an RTT of $33$ms, while a concurrent one is downloading a similar file from the server in Bangalore. To ensure these flows share some bottleneck link we route this latter through the cloud ($\rs$ and $\rc$), without splitting it. The resulting RTT for this route is therefore $274$ms. The two HTTP requests from the client are sent simultaneously. As expected, the flow with the shorter RTT hogs the bottleneck link, and the longer-RTT flow ramps up only at the end of the shorter flow.
    }
    \label{fig:direct-vs-nat}
\end{figure}

\begin{figure}[!t]
  \centering
    \includegraphics[width=0.46\textwidth,trim=20mm 25mm 25mm 20mm,clip]{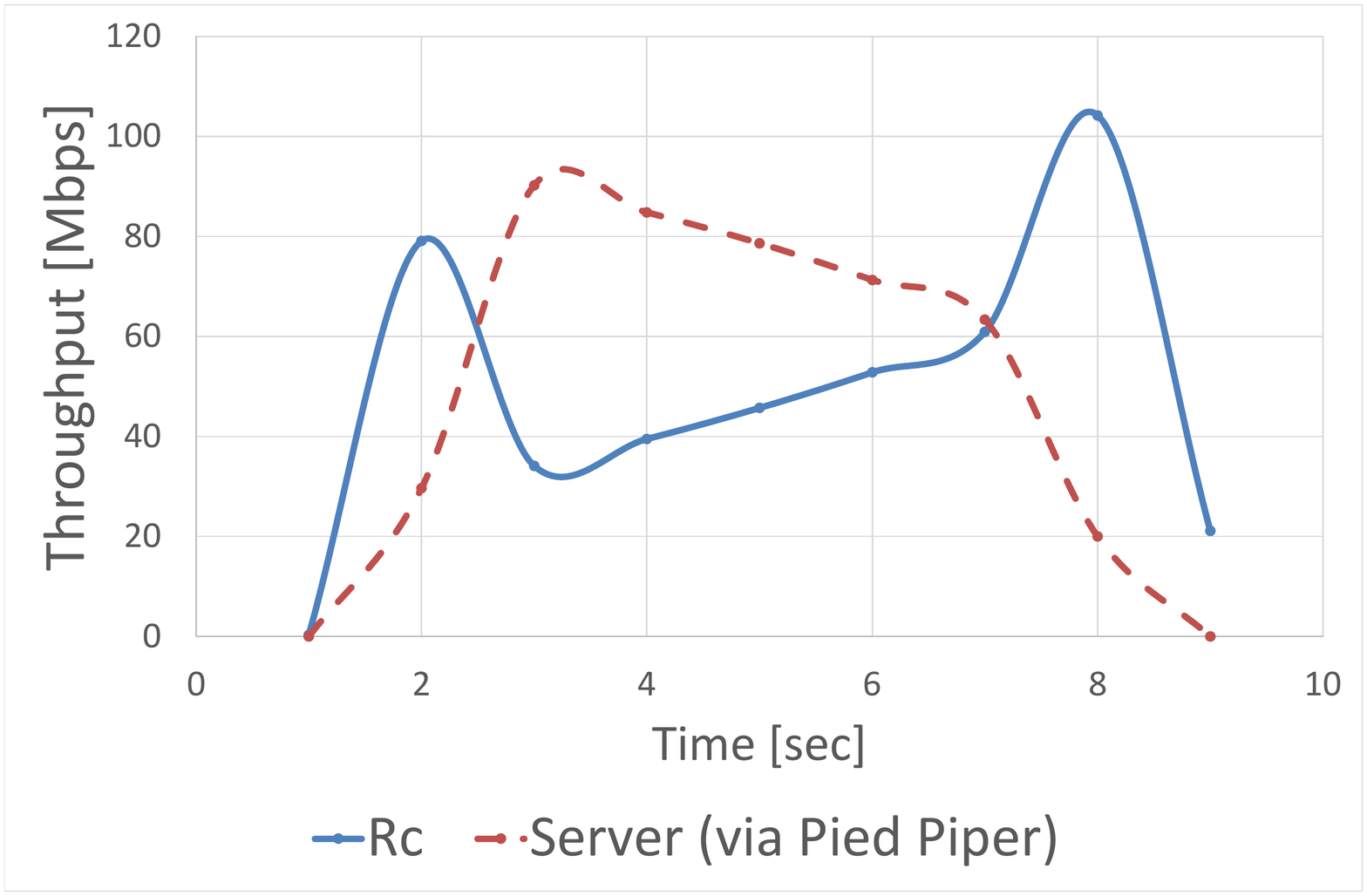}
    \caption{Fairness between a flow using \name to download a 50 MB file from the server and a competing flow (destined to the same client) downloading a similar 50 MB file from $\rc$ (the relay used by \name as well). 
    }
    \label{fig:direct-vs-piper}
\end{figure}

However, OCD does have a noticeable effect on TCP fairness. \name, in fact, interestingly \textit{improves} the fairness between long-haul TCP flows and their shorter competitors.
It is well known that TCP Cubic (as well as other TCP congestion control algorithms) is unfair towards large-RTT flows when they share a bottleneck connection with short RTT flows \cite{kelly2001mathematical}. Indeed, we conducted such an experiment, using our infrastructure to route the long-haul flow through our relays, without otherwise modifying the flow, and have it share the $(\mbox{client}\Longleftrightarrow\rc)$ route with a shorter flow downloading directly from $\rc$.
\autoref{fig:direct-vs-nat} presents the throughput of each of the flows over time (averaged over a window of 1 sec). We clearly see that the flow with the shorter RTT actually pushes out the longer-RTT flow. When the same experiment is run, this time enabling \name on the large-RTT flow (\autoref{fig:direct-vs-piper}), the improvement in the throughput of the long-haul flow is apparent. This time, both flows get a roughly equal share of the bottleneck link.
 
\section{Related Work}\label{sec:related-work}

\T{Internet overlays.} The idea of overlay networking on top of the Internet dates back almost two decades~\cite{old-overlay-1, old-overlay-2, RON}. Many studies established that the default BGP path between two end-points is often inferior to an alternate path traversing an overlay network~\cite{old-overlay-1, old-overlay-2, RON, akamai-2, akamai-3, akamai-4}.

\T{Cloud overlays.} Recently, overlay networking through the \emph{cloud} has received attention from both researchers (e.g.,~\cite{CRONets}) and practitioners (e.g.,~\cite{teridion}). Work along these lines often explicitly or implicitly reflects the understanding that the Internet path to/from the cloud is the bottleneck whereas the cloud provides ample network capacity~\cite{le2016understanding,jeyakumar2012eyeq}.

CRONets~\cite{CRONets} leverages a \emph{single} cloud relay with TCP splitting for data delivery. Our experimentation with a broad spectrum of routing and congestion control schemes for cloudified data delivery, suggests that cloud overlays can provide significantly higher performance than that achievable using only a single relay.

\T{Cloud-based content delivery.} The use of an HTTP proxy~\cite{cgn2017} in static AWS locations was shown to reduce webpage load times by half, when accessing remote websites. Cloud-based mobile browsing~\cite{zhao2011reducing,wang2013accelerating} is a common technique for reducing cellular costs and decreasing download times.

\T{Cloud connectivity and performance.} \cite{one-hop} shows that some 60\% of end-user prefixes are within one-AS-hop from GCP (Google Cloud Platform). A different study~\cite{cgn2017} established that AWS has at least one node within a median RTT of merely 4.8~ms from servers hosting the top 10k most popular Web sites. \cite{unusual} shows that modern clouds have unusual internal routing structures that are hard to infer and CLAudit \cite{multidimensional} points out the increasing reliability of cloud latency. 

\T{TCP Split.} Several papers considered TCP split to better TCP performance, \eg overcome different link characteristics~ \cite{Kopparty2002} (wired and wireless), compensate for very long RTTs in satellite links \cite{luglio2004}, and reduce search query latency~\cite{pathak2010measuring}. 
Pucha and Hu explore TCP overlay \cite{pucha2005overlay, pucha2005slot}. 

Miniproxy~\cite{siracusano2016miniproxy} implements split TCP proxy in a virtualized environment. In contrast to our work, miniproxy utilizes minikernels~\cite{minikernel} and a modified lwip \cite{lwip} instead of \sockets. The only previous work that we know of, that implements a Split TCP proxy in the kernel is~\cite{kernelsplit}. This work simulates a split connection by spoofing ACKs and forwarding the packets, working packet by packet rather than with byte streams. We, in contrast, use standard \sockets and manage each side of the connection separately. This allows us to be flexible and easily experiment with different optimizations.

\section{Conclusion}\label{sec:conclusion}
We initiated the systematic exploration of cloudified data delivery and presented initial results suggesting that optimization of performance should be viewed through the congestion-control lens.  We view cloudified data delivery as a promising direction for overcoming the inefficiencies of today's routing (BGP) and congestion control (TCP) protocols. We thus argue that tapping the full potential of this approach is of great importance.

\bibliographystyle{acm}
\bibliography{mybib,overlaybib,sigcomm18}

\end{document}